\documentclass{iopart}

\usepackage[dvips]{graphics}
\usepackage{graphicx}
\usepackage{psfrag}
\usepackage{subfigure}

\renewcommand{\b}{\mathbf}

\begin{document}

\title{Nonlocal spectral properties of disordered alloys}

\author{G.~M.~Batt and D.~A.~Rowlands}
\address{H.H.~Wills Physics Laboratory, University of Bristol, Bristol BS8 1TL, U.K.}

\begin{abstract}

A general method is proposed for calculating a fully $\b{k}$-dependent, continuous, and causal spectral function $A(\b{k},E)$ within the
recently introduced nonlocal version of the coherent-potential approximation (NLCPA). The method involves the combination of both periodic
and anti-periodic solutions to the associated cluster problem and also leads to correct bulk quantities for small cluster sizes. We
illustrate the method by investigating the Fermi surface of a two-dimensional alloy. Dramatically, we find a smeared electronic topological
transition not predicted by the conventional CPA.

\end{abstract}


\submitto{\JPCM}

\section{Introduction}

The coherent-potential approximation (CPA)~\cite{Soven1} is a successful and widely used theory of dealing with disordered solids. However,
the CPA does have a serious limitation in that it is only a single-site mean-field theory of disorder and hence unable to treat multi-site
correlations in the ensemble of disorder configurations. However, this long-standing problem has recently been addressed by the development
of the nonlocal coherent-potential approximation (NLCPA)~\cite{Jarrell1}, a cluster mean-field theory which provides systematic corrections
to the CPA. It was introduced as the static version of the dynamical cluster approximation (DCA)~\cite{Hettler1,Hettler2} used to describe
short-range correlations in strongly-correlated electron systems, and has been implemented both for model
Hamiltonians~\cite{Jarrell1,Moradian2,Rowlands2} and within first-principles density functional theory~\cite{Rowlands3,Rowlands4,Rowlands5}.
Notably, in contrast to earlier cluster theories~\cite{Tsukada1} and supercell calculations, the NLCPA preserves the full translational
symmetry of the underlying lattice. Significantly, this highly desirable attribute means the NLCPA could be used to study nonlocal effects on
spectral properties of disordered systems. Indeed, the spectral function $A_B(\b{k},E)$ for metallic alloys is of general interest because
its peaks define the Fermi surface. For an ordered system these peaks are delta functions, however for a substitutionally disordered alloy
they have finite heights and widths $\xi^{-1}$, where $\xi$ is the coherence length at the Fermi energy $E_F$. The shapes of such Fermi
surfaces can be determined from two-dimensional Angular Correlation of (Positron) Annihilation Radiation (2D-ACAR)
experiments~\cite{Wilkinson1}, and, since they drive various ordering processes~\cite{Gyorffy1} and are subject to observable topological
changes~\cite{Varlamov1,Kordumova1}, their theoretical description is an important area in alloy physics. One of the main objectives of this
paper is to introduce a method for calculating the spectral function within the NLCPA and to demonstrate its use by explicit calculations for
the Fermi surface of a tight-binding model alloy.

Recall that for such a tight-binding model alloy, the configurational average over all possible disorder configurations is represented by a
translationally-invariant effective medium characterized by the exact self energy $\Sigma(\b{k})$. By mapping to a self-consistent
single-site impurity problem, the conventional CPA neglects correlations beyond a single site and hence yields a single-site self-energy
$\Sigma$ that has no dependence on momentum $\b{k}$. The idea of the NLCPA is to systematically restore this neglected $\b{k}$-dependence by
replacing the conventional single-site impurity with an ensemble of impurity cluster configurations, thus providing systematic corrections to
the CPA as the cluster size increases. Unlike previous cluster theories, the NLCPA preserves translational-invariance by imposing Born-von
Karman (periodic) boundary conditions on the cluster. However, a consequence of this is that the resulting effective medium is characterized
by a self-energy $\Sigma(\b{K})$ which is a \emph{step function} in $\b{k}$-space, with the number of steps centred at the points $\b{K}$
equal to the number of sites in the cluster. This naturally leads to a spectral function with \emph{discontinuities} at the jumps, which in
turn leads for example to an unphysical Fermi surface.

Furthermore, it has recently been explicitly shown~\cite{Rowlands2} that the NLCPA results are different if
\emph{anti-periodic}~\cite{Hettler2,Rowlands2} Born-von Karman boundary conditions are imposed on the same cluster instead of the
conventional periodic boundary conditions. The periodic and anti-periodic results only become equivalent as the cluster size $N_c$ is
increased to some critical value which depends on the model in question. In fact, below this critical value of $N_c$ it has been explicitly
shown that the periodic and anti-periodic solutions on their own are unphysical due to the discontinuities, i.e.~not sufficient to yield a
meaningful density of states~\cite{Rowlands2}.

In this paper we introduce a new method for implementing the NLCPA which resolves all the issues above. It involves the combination of both
periodic and anti-periodic solutions to the NLCPA and should be used for cluster sizes smaller than the critical value. It yields a fully
$\b{k}$-dependent, continuous, and causal spectral function $A(\b{k},E)$, together with a meaningful density of states.

The outline of this paper is as follows. First, we briefly summarise the idea of the NLCPA in section \ref{summary}. Then, in section
\ref{idea} the concept of periodic and anti-periodic Born-von Karman boundary conditions is described in detail, and the key idea of the new
method for implementing the NLCPA is introduced. The strict criteria that the new method must satisfy are outlined in \ref{conditions}, and
the formalism is detailed in section \ref{choice}. To test the validity of the new method, section \ref{validity} describes an application to
a simple 1D tight-binding model in order to compare with known results. Following this, we apply the method to a 2D tight-binding model in
section \ref{fermi2d} and investigate the Fermi surface. Finally we conclude in section \ref{conclusions}.

\section{Formalism}\label{formalism}

\subsection{Brief summary of the NLCPA}\label{summary}

Consider a general tight-binding model binary alloy defined by a hopping amplitude $W_{ij}$ and site energies $V_A$ and $V_B$. In terms of
the site occupation numbers $\xi_i$ which take on values of 0 and 1 depending on whether the site is occupied by an $A$-atom or a $B$-atom,
the usual Green's function $G_{ij}(E)$ satisfies the equation
\begin{equation}
 \sum_{ij}\left[(E-V_i)\delta_{ij}-W_{ij}\right]G_{ij}=\delta_{ij}
\end{equation}
where $V_i=\xi_i V_A+(1-\xi_i) V_B$ and $E$ is the energy.
 We are
interested in the average of $G_{ij}(E)$ over all possible disorder
configurations $\{\xi_i\}$. Here it is useful to introduce the
concept of an effective medium, and rewrite the average Green's
function in terms of the exact self-energy $\Sigma_{ij}$ which is
translationally-invariant. In essence it replaces the substitutional
disorder at each lattice site. By applying a Fourier transform to
such translationally-invariant quantities we obtain
\begin{equation}\label{Gexact}
  \overline{G}(\b{k})=\left(E-W(\b{k})-\Sigma(\b{k})\right)^{-1}
\end{equation}
where $a$ is the lattice constant and $W(\b{k})$, $\Sigma(\b{k})$
are the hopping and exact self-energy respectively in reciprocal
space. In the interests of clarity, the energy dependence of all
Green's functions on energy will be implicitly defined as in
(\ref{Gexact}) for the remainder of this paper.

In the single-site CPA the self-energy is $\b{k}$-independent and the purpose of the NLCPA is to systematically restore this neglected
$\b{k}$-dependence. As explained in \cite{Jarrell1,Rowlands2} the idea is to coarse-grain the function $\Sigma(\b{k})$ over a set of $N_c$
tiles which divide up the Brillouin zone of the underlying lattice. If these tiles are centred at a set of $N_c$ points $\{\b{K}_n\}$ then
this results in a function $\Sigma(\b{K}_n)$ which is a step function in $\b{k}$-space. The NLCPA is able to determine the function
$\Sigma(\b{K}_n)$ by mapping to a self-consistent embedded cluster problem, where the cluster of $N_c$ sites has Born-von Karman (periodic)
boundary conditions imposed. This means the real-space cluster sites $\{\b{R}_I\}$ and corresponding \emph{cluster momenta} $\{\b{K}_n\}$
are related by the Fourier transform
\begin{equation}\label{Eqn:nltransform}
 \frac{1}{N_c}\sum\limits_{\b{K}_n}\sum\limits_{IJ}e^{i\b{K}_n\cdot(\b{R}_I-\b{R}_J)}=1
\end{equation}
where $n=1,..,N_c$. This relation is key to constructing an
algorithm to determine the medium (see \cite{Jarrell1} for full
details). After convergence of the algorithm, the NLCPA
approximation to the exact average Green's function (\ref{Gexact})
is given by
\begin{equation}\label{Gapprox}
 \overline{G}(\b{k})=\left(E-W(\b{k})-\Sigma(\b{K}_n)\right)^{-1}
\end{equation}
where $\Sigma(\b{K}_n)$ takes its appropriate value within each reciprocal-space tile $n$. Clearly the effect of coarse-graining the
self-energy has been to reduce its range in real space (although physics on a longer length scale is treated at the mean-field level),
however translational invariance is always preserved.

Unfortunately, as mentioned in the introduction, the fact that $\Sigma(\b{K}_n)$ takes its appropriate value within each reciprocal-space
tile in (\ref{Gapprox}) means that discontinuities have been introduced at the tile boundaries, leading for example to an unphysical Fermi
surface.

\subsection{Idea of the new method}\label{idea}

In order to remove the discontinuities in $\overline{G}(\b{k})$ described above, an idea would be simply to interpolate the self-energy
$\Sigma(\b{K}_n)$ to a smooth function $\Sigma(\b{k})$ in such a way as to preserve the number of states. However, the problem with such an
approach is that new information (e.g.~structure in the DOS) would be inferred. A second problem is that of preserving causality. For example
previous attempts have been made (within the DCA)~\cite{Maier3} to interpolate the self-energy via schemes such as the maximum entropy method
(MEM). However, although the MEM is guaranteed to be causal, it can only be used to add a finite number of points to a data distribution,
after which it adds spurious structure. One must use a secondary interpolation to fill in the gaps left by the MEM e.g.~Akima spline as used
in the DCA~\cite{Maier3}. However such spline interpolations are well known to overshoot the interpolation in some
circumstances~\cite{Fried1}, and particulary when the data set used for the interpolation is rapidly varying the spline may violate
causality. To overcome this, more data points are required and hence a bigger cluster size, but one is still not ensured a causal
interpolation for all cases. Furthermore, in the case of first-principles alloy calculations~\cite{Rowlands3,Rowlands4,Rowlands5} we are
particularly interested in large deviations from the CPA in $\b{k}$-space where the self-energy will vary rapidly since only small clusters
are computationally viable at the present time.

\begin{figure}[!]
 \begin{center}
\scalebox{0.65}{\includegraphics{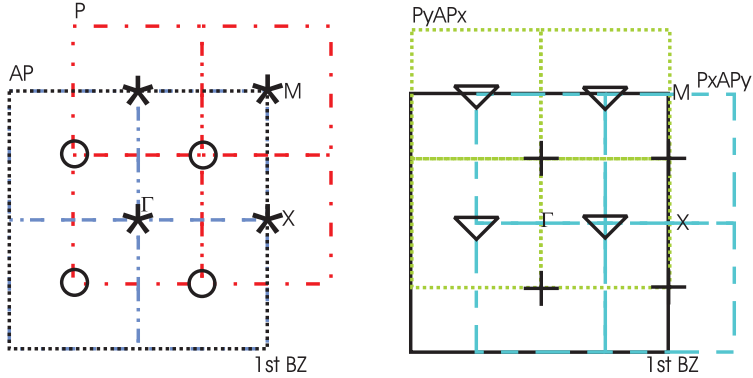}}
   \caption{(colour online) Shown \textbf{(left) }are the sets of cluster momenta $\{\b{K}_n^P\}$ (denoted by asterisk) and $\{\b{K}_n^{AP}\}$ (denoted by
     circles) centred at their corresponding tiles. Shown \textbf{(right)} are the sets $\{\b{K}_n^{P_x{AP}_y}\}$ (denoted by plus) and
     $\{\b{K}_n^{P_y{AP}_x}\}$ (denoted by trangles) centred at their corresponding tiles. Note that $\Gamma$ is situated at $(0,0)$,
     $X$ is situated at $(0,\pm\frac{\pi}{a})$ and $(\pm\frac{\pi}{a},0)$, and $M$ is situated at $(\pm\frac{\pi}{a},\pm\frac{\pi}{a})$.}
         \label{fig1}
 \end{center}
\end{figure}

In this paper a different approach to the problem is introduced. First recall that in previous NLCPA calculations to date, periodic Born-von
Karman boundary conditions have been applied to the cluster. This essentially means reducing the size of a conventional lattice with Born-von
Karman boundary conditions to contain only a cluster of $N_c$ sites, so that the edges of the cluster map round to the other end in each
dimension e.g.
\begin{equation}
 \overline{G}\left({R_x+L_x,R_y+L_y,R_z+L_z}\right)=\overline{G}\left({R_x,R_y,R_z}\right)
\end{equation}
for the cluster Green's function, where $L_i$ is the length of
the cluster along direction $i$. Since the lattice constant is
unchanged, the boundaries of the BZ will remain the same, however it
will now contain only $N_c$ evenly spaced $\b{k}$ points referred to
as the set of cluster momenta $\{\b{K}_n\}$. Thus the conventional
lattice Fourier transform used in the $N_c\rightarrow\infty$ limit
reduces to the cluster Fourier transform given by
(\ref{Eqn:nltransform}). However, as mentioned in the introduction,
for a given cluster there are another set of cluster momenta
$\{\b{K}_n\}$ satisfying (\ref{Eqn:nltransform}) corresponding to
\emph{anti-periodic}~\cite{Hettler2,Rowlands2} Born-von Karman
boundary conditions so that quantities such as the cluster Green's
function at the edges of the cluster map back to minus the value at
the other end i.e.
\begin{equation}
 \overline{G}\left({R_x+L_x,R_y+L_y,R_z+L_z}\right)=-\overline{G}\left({R_x,R_y,R_z}\right)
\end{equation}
As an example, the periodic and anti-periodic set of cluster momenta for a 2D square lattice with $N_c=4$ are shown in figure
\ref{fig1}, labelled as $\{\b{K}_n^P\}$ and $\{\b{K}_n^{AP}\}$ respectively. Permutations of periodic and anti-periodic conditions along
each axis are also possible, for example periodic along the x-axis and anti-periodic along the y-axis and vice-versa. These are shown in
figure \ref{fig1} for the same example and are labelled as $\{\b{K}_n^{P_x{AP}_y}\}$ and $\{\b{K}_n^{P_y{AP}_x}\}$ respectively.

Next, recall that it has recently been explicitly shown~\cite{Rowlands2} that for a given cluster the NLCPA results change with respect to
the choice of $\{\b{K}_n\}$. Any theory with a choice of boundary condition must be invariant in its predictions of bulk
properties~\cite{Ashcroft1} and hence at present the NLCPA can only be perceived as a unique theory above some critical value of $N_c$, which
depends on the model in question. Moreover, below this critical value of $N_c$ it has been explicitly shown that the periodic and
anti-periodic DOS results on their own are not meaningful due to the discontinuities~\cite{Rowlands2}. The main idea of the method introduced
in this paper is to appropriately combine the results obtained with the various boundary conditions to give a unique and meaningful result
for any given cluster.

First note that, significantly, the set of $\b{K}_n$ points
associated with each boundary condition is shifted compared to the
others so that no overlap occurs. This can be clearly seen in figure
\ref{fig1} for the 2D square lattice example. Next, let us label the
Green's function (given by (\ref{Gapprox})) obtained using a
particular boundary condition by $\overline{G}_{i}(\b{k})$ and its
corresponding set of cluster momenta by $\{\b{K}_n\}_{i}$, where
$i=P,AP,P_x{AP}_yP_z..$ etc. Note that on account of the
coarse-graining procedure outlined in section \ref{summary}, by
construction $\overline{G}_{i}(\b{k})$ is a better approximation to
the exact result in the region of reciprocal space close to each of
the points $\{\b{K}_n\}_{i}$ (in fact other
$\overline{G}_{j}(\b{k})$ with $j{\neq}i$ have discontinuities at
$\{\b{K}_n\}_{i}$). This suggests that one should construct a new
Green's function $\overline{G}_{M}(\b{k})$ which follows each
$\overline{G}_{i}(\b{k})$ close to each of its corresponding set of
cluster momenta $\{\b{K}_n\}_{i}$ only, smoothly joining up with the
other $\overline{G}_{i}(\b{k})$ close to their set of
$\{\b{K}_n\}_{i}$. Such a combined solution
$\overline{G}_{M}(\b{k})$ is guaranteed to be causal since it will
always lie between causal extremes $\overline{G}_{i}(\b{k})$.
Furthermore, the problem of discontinuities is automatically removed
since no solution would be followed where it has a discontinuity.

In addition to yielding a new mixed Green's function
$\overline{G}_{M}(\b{k})$ which is fully $\b{k}$-dependent with no
discontinuities, observe that the method effectively increases the
number of tiles dividing up the BZ. This leads to a better sampling
of the BZ and is thus expected to yield meaningful bulk quantities
such as density of states. The strict criteria that the method must
satisfy are outlined in the next section.

\subsection{Formalism: Introduction of continuous $\b{k}$-dependence}\label{conditions}

In section \ref{idea} it was proposed that a new unique solution
(Green's function) may be constructed by combining solutions
associated with the different choices of $\{\b{K}_n\}$. We propose
the following criteria that the combined solution
$\overline{G}_M(\b{k})$ must obey:
\begin{enumerate}
  \item  Follow the individual solutions through their associated $\b{K}_n$-points (hence yielding a meaningful DOS).
  \item  Be based on a well founded mathematical principle such as the variational principle.
  \item  Yield a spectral function $A_B(\b{k})$ with no discontinuities.
  \item  Be causal everywhere in ${\b{k}}$-space.
  \item  Conserve the total number of electron states $N$.
  \item  Retain the point-group symmetry of the lattice.
\end{enumerate}
Here we propose that a general mixed solution
$\overline{G}_M(\b{k})$ satisfying all the above criteria may be
written in the form
\begin{equation}\label{mix}
 \overline{G}_M(\b{k}) = \sum\limits_i\gamma_i\left( \b{k} \right)\overline{G}_i
 \left(\b{k}\right)
\end{equation}
where $\gamma_i({\b{k}})$ is the mixing parameter of the solution
$\overline{G}_i(\b{k})$ defined by (\ref{Gapprox}), with $i=P,AP..$
etc. This mixing parameter is generally $\b{k}$-dependent with range
\begin{equation}\label{range}
 0\leq\gamma_i(\b{k})\leq1
\end{equation}
and must satisfy
\begin{equation}\label{Eqn:Sumgamma}
 \sum\limits_i\gamma_i\left(\b{k}\right)=1
\end{equation}
For example, for the case of a 2D square lattice the new Green's function may be written as
\begin{eqnarray}
  {\overline{G}_M}(\b{k})&=&\gamma_P({\b{k}})\overline{G}_P({\b{k}})+\gamma_{AP}({\b{k}})\overline{G}_{AP}({\b k}) \nonumber\\
                         &+&\gamma_{P_y{AP}_x}({\b{k}})\overline{G}_{P_y{AP}_x}(\b{k})+\gamma_{P_x{AP}_y}({\b{k}})\overline{G}_{P_x{AP}_y}(\b{k})
\end{eqnarray}
where each $\gamma_i(\b{k})$ is to be determined. The new Green's
function $\overline{G}_M(\b{k})$ will obviously yield a new set of
bulk properties and a unique DOS with its own Fermi energy $E_F$.
The spectral function can then be defined in terms of
$\overline{G}_M(\b{k})$ by
\begin{equation}\label{eqn:spect}
 A_B^M(\b{k})=-\frac{1}{\pi}Im\,{\overline{G}_M(\b{k})}
\end{equation}
Using $\overline{G}_M(\b{k})$, a new self-energy $\Sigma_M({\b{k}})$
may also be defined from Dyson's equation
\begin{equation}
 \Sigma_M({\b{k}})=G_0^{-1}({\b{k}})-\overline{G}_M^{-1}({\b{k}})
\end{equation}
In \ref{causality} it is shown that $\overline{G}_M(\b{k})$
satisfies criteria (iv) and (v) above for any set of mixing
parameters $\gamma_i(\b{k})$ satisfying (\ref{range}) and
(\ref{Eqn:Sumgamma}). Next, in section \ref{choice} it is shown
generally how to choose a specific set of $\gamma_i({\b{k}})$ so
that the remaining criteria are satisfied.

\subsection{Formalism: Choice of $\gamma(\b{k})$}\label{choice}

Given the functional requirements described in section
\ref{conditions} that the $\gamma_i(\b{k})$ must satisfy, one may
narrow down a choice of suitable functions. First,
$\overline{G}_M(\b{k})$ must follow each $\overline{G}_i(\b{k})$ as
close as possible to its associated set of cluster momenta
${\{\b{K}_n\}}_i$. As explained in section \ref{idea} and is clear
from figure \ref{fig1}, this enables one to remove all
discontinuities since no solution will be followed at its
corresponding tile boundary where the discontinuities occur.
Specifically, $\overline{G}_i(\b{k})$ must equal zero at the tile
boundaries associated with its $\{\b{K}_n\}_i$ and so each
$\gamma_i(\b{k})$ must be a function that is equal to 1 in the tile
centres $\{\b{K}_n\}_i$ and smoothly change to a value of
$\gamma_i(\b{k})=0$ on the tile boundaries. Thus the discontinuities
in $\overline{G}_i(\b{k})$ are removed by cancellation with
$\gamma_i(\b{k})$. $\overline{G}_i(\b{k})$ must also be a periodic
function in $\b{k}$-space with the same periodicity as
$\{\b{K}_n\}_i$. Therefore $\gamma_i(\b{k})$ should be expandable in
a Fourier series made up of terms from (\ref{Eqn:nltransform}). We
also want to base our choice on a mathematical principle that gives
a unique solution.

Generally for this system we can define a potential $\mu$ that is
proportional to the density of states (DOS)
\begin{equation}\label{pot}
\mu \propto \int \sum_i \overline{G}_i({\bf {k}})\gamma_i({\bf {k}})
d{\bf {k}}
\end{equation}
For example if a power of the system energy $E^n$ is used as the proportionality factor one can immediately see that the moments (e.g. band
energy) will be directly affected by the choice of $\gamma_i(\b{k})$. Hence, the choice of $\gamma_i(\b{k})$ cannot be considered arbitrary, one must aim
to minimize $\mu$ through $\gamma_i(\b{k})$. At any point in ${\bf {k}}$-space there will be an associated potential
\begin{equation}\label{ga}
\mu_{\gamma_i}(\b{k}) \propto -\gamma_i(\b{k}) .
\end{equation}
between $\overline{G}_M(\b{k})$ and $\overline{G}_i(\b{k})$. Each
$\mu_{\gamma_i}(\b{k})$ has well defined limits, since at its
${\b{K}_n}$-points it should be at a minimum, where the NLCPA
coarse-grained $\overline{G}_i({\b{K}_n})$ are considered best. Also
it will be a maximum on the tile boundary where
$\overline{G}_i({\b{K}_n})$ is undefined and is at its worst. From
section \ref{conditions}, $\gamma_i(\b{k})$ is independent of the
value $\overline{G}_i(\b{k})$, and hence our mixing potential
(\ref{ga}) will have the same proportionality factor for each
$\gamma_i(\b{k})$ which we can denote by the value $\tau$. The total
potential of $\gamma_i(\b{k})$ is hence given by
\begin{equation}\label{gam}
\mu_\gamma(\b{k}) = \tau\sum_i -\gamma_i(\b{k})
\end{equation}
Notice that $\mu_\gamma(\b{k})$ at the $\b{K}_n$ points will be
solely attributed to the associated $\overline{G}_i(\b{k})$ at that
point and vice versa on its tile boundary and we have assumed no
functional form for $\gamma_i(\b{k})$. With the boundary conditions
on how each $\gamma_i(\b{k})$ must behave well defined we seek to
minimize (\ref{gam}). To solve this problem one must notice that
(\ref{gam}) can be mapped directly to the potential of a static
membrane~\cite{Courant1} described by the general solution of the
wave equation $u(\b{k})$, related to $\gamma_i(\b{k})$ by
\begin{equation}\label{diffa}
1/2\nabla u(\b{k}).\nabla u^*(\b{k}) \propto \sum_i\gamma_i(\b{k})
\end{equation}
since $1/2\nabla u(\b{k}).\nabla u^*(\b{k})$ to a first order is proportional to that of a potential stored in a static membrane~\cite{Courant1}. Hence the
sum of $\gamma_i(\b{k})$ is equal to the differential surface area of the membrane mapped out by $u(\b{k})$. We wish to minimize the total potential
$\sum_i \mu_{\gamma_i}(\b{k})$ while treating all $\gamma_i(\b{k})$ equally. Minimization of (\ref{diffa}) gives the well known Euler (Laplace) equation
describing wave motion
\begin{equation}
\nabla^2u(\b{k})=0
\end{equation}
that is solved generally for a membrane of the same dimensions as the tile for the problem i.e.~we only allow waves that can propagate
according the boundary conditions of $\gamma_i(\b{k})$.

\subsubsection*{Example: 2D square lattice}

As an example, for the 2D square lattice $u$ has the general form
\begin{equation}\label{gen}
u({\bf{k}})= A\exp(i(L_x/2)k_x)\exp(i(L_y/2)k_y)
\end{equation}
where the origin is defined at the $\Gamma$-point in the first Brillouin Zone, $L_x$ and $L_y$ are the lengths of the cluster (principle
vectors) along the $x$ and $y$ directions respectively, and $A$ is a complex constant. Substituting (\ref{gen}) into (\ref{diffa}) gives
\begin{eqnarray}\label{moduex}
 \sum_i\gamma_i(\b{k})& \propto &  AA^*((L_x/2)^2+(L_y/2)^2)\times \nonumber\\
 &   & [\cos^2((L_x/2)k_x)+\sin^2((L_x/2)k_x)]\times \nonumber\\
 &   & [\cos^2((L_y/2)k_y)+\sin^2((L_y/2)k_y)] \nonumber\\
 & = & AA^*((L_x/2)^2+(L_y/2)^2)\times \nonumber\\
 &   & [\cos^2((L_x/2)k_x)\cos^2((L_y/2)k_y) \nonumber\\
 &   &  +\sin^2((L_x/2)k_x)\sin^2((L_y/2)k_y) \nonumber\\
 &   & +\cos^2((L_x/2)k_x)\sin^2((L_y/2)k_y)  \nonumber\\
 &   &  + \cos^2((L_y/2)k_y)\sin^2((L_x/2)k_x)] \nonumber\\
 & = & AA^*((L_x/2)^2+(L_y/2)^2)
\end{eqnarray}
where $\gamma_i(\b{k})$ can now be determined by comparison of
(\ref{moduex}) and (\ref{Eqn:Sumgamma}) to yield
\begin{eqnarray}
 \gamma_P({\bf {k}})           &=& (\cos((L_x/2)k_x)\cos((L_y/2)k_y))^2 \nonumber\\
 \gamma_{AP}({\bf {k}})        &=& (\sin((L_x/2)k_x)\sin((L_y/2)k_y))^2 \nonumber\\
 \gamma_{P_y{AP}_x}({\bf {k}}) &=& (\sin((L_x/2)k_x)\cos((L_y/2)k_y))^2 \nonumber\\
 \gamma_{P_x{AP}_y}({\bf {k}}) &=& (\cos((L_x/2)k_x)\sin((L_y/2)k_y))^2.
\end{eqnarray}
Each $\gamma_i(\b{k})$ can be seen to uphold their respective
boundary conditions given in section \ref{conditions}. Each solution
to $\overline{G}_i(\b{k})$ is thus treated equally while removing
all discontinues and minimizing the potential $\mu_\gamma(\b{k})$,
and that of $\mu$ given the restrictions on $\gamma_i(\b{k})$.

Note that each $\gamma_i(\b{k})$ smoothly varies between a value of
$1$ at $\{\b{K}_n\}_i$ and $0$ at the associated tile boundary and
has the same periodicity as the tiling. Also, because the sum over
$\gamma_i(\b{k})$ in (\ref{Eqn:Sumgamma}) is always satisfied even
at the tile boundaries one can see that all discontinuities are
removed because the solution remains continuous and bounded  (lies
in the range between the smallest and largest solutions of
$\overline{G}_i(\b{k})$ given by (\ref{range})). At the
$\b{k}$-points where there are no $\{\b{K}_n\}_i$ the method will
take a mix of $\overline{G}_i(\b{k})$ with weights corresponding to
the locality of their $\{\b{K}_n\}_i$ points. It is also worth
mentioning the limiting cases of $N_c=1$ and $N_c\rightarrow\infty$.
When $N_c=1$ all $\overline{G}_i(\b{k})$ will equal the same CPA
value. This is because there will only be one tile (which
encompasses the whole Brillouin Zone) where $\gamma_i(\b{k})$ always
satisfies (\ref{Eqn:Sumgamma}), thus $\overline{G}_M(\b{k})$ will
also equal the CPA value. When $N_c$ approaches infinity all
$\overline{G}_i(\b{k})$ will approach the exact solution for
$\overline{G}(\b{k})$ given by (\ref{Gexact}) hence so must
$\overline{G}_M(\b{k})$ because (\ref{Eqn:Sumgamma}) bounds them
within the range of $\overline{G}_i(\b{k})$. Also note that
(\ref{gen}) can be extended to 3-dimensions with the inclusion of a
${\b{k}_z}$ term of the same form of the other dimensions, where all
mathematical properties of $\gamma_i(\b{k})$ shown for the 2D case
will hold.

\section{Results}\label{results}

\subsection{Validity of the new method}\label{validity}

\begin{figure}[!]
 \begin{center}
 \begin{tabular}{ccc}
   \psfrag{ECM}[B1][B1][2][0]{ECM}\scalebox{0.33}{\includegraphics{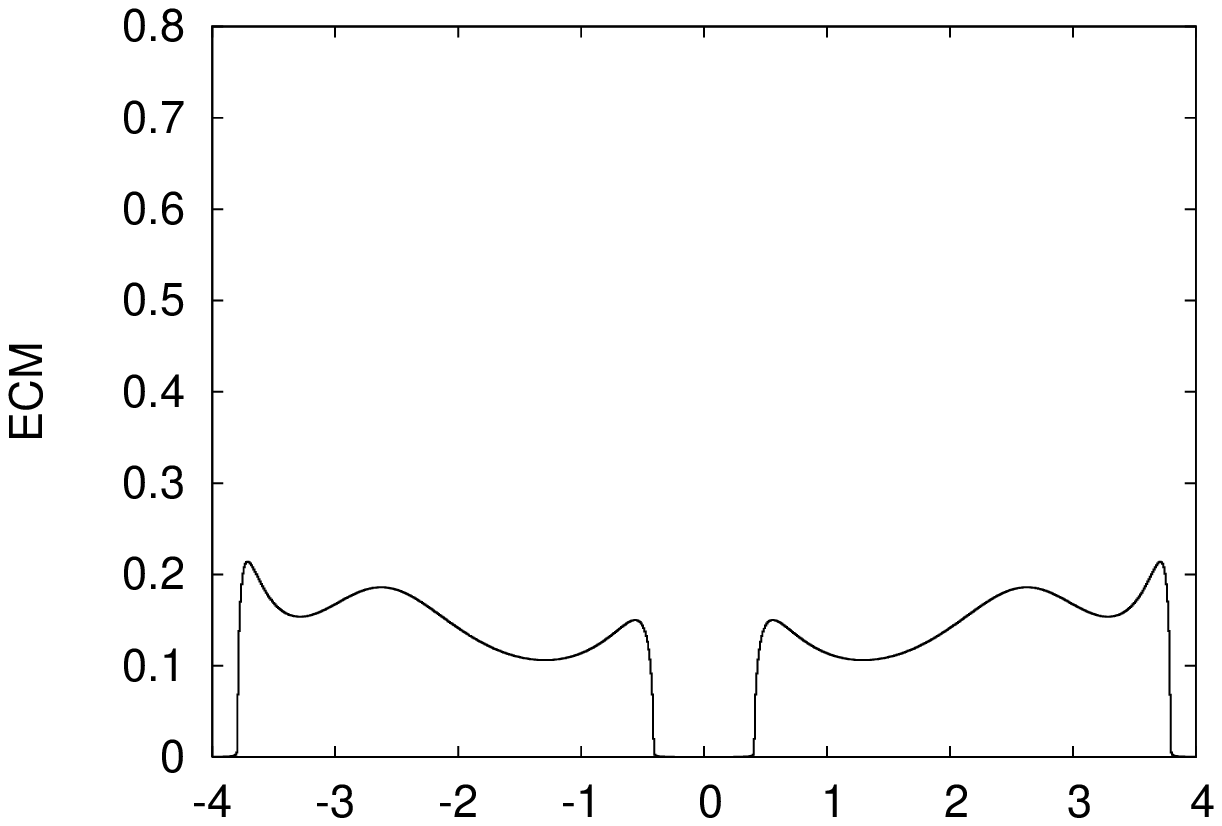}}
 & \scalebox{0.33}{\includegraphics{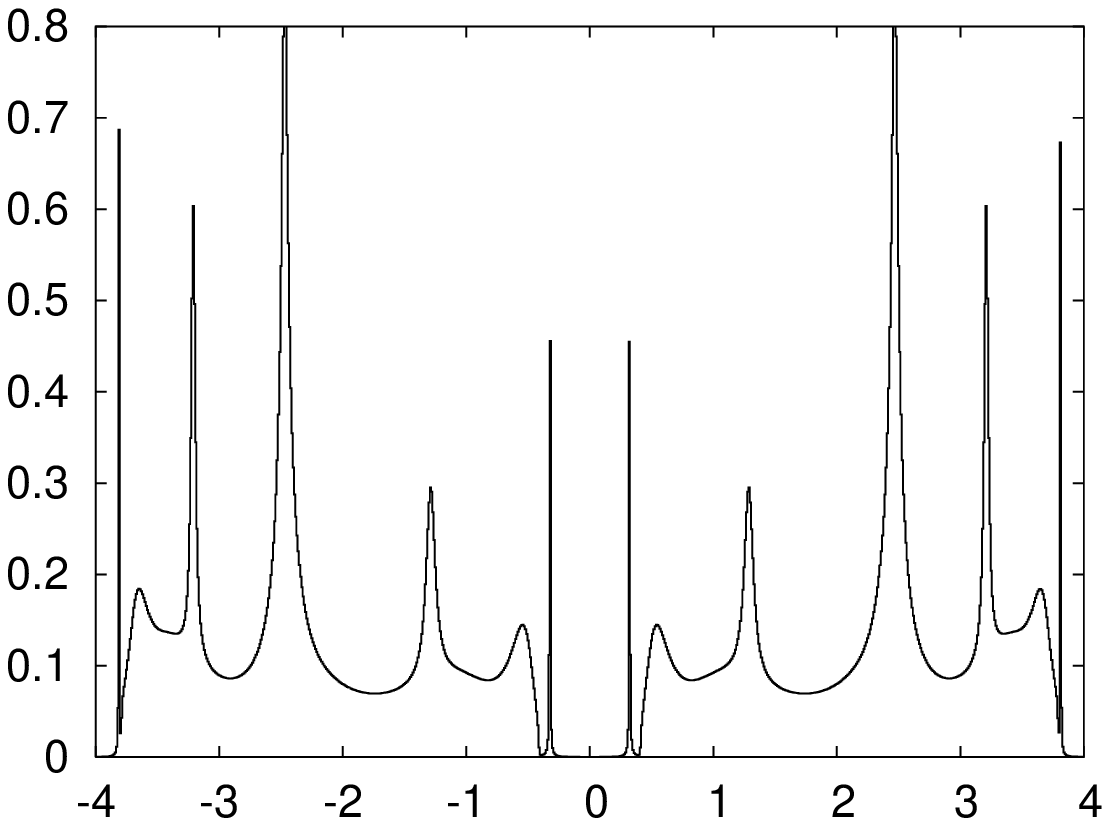}}
 & \scalebox{0.33}{\includegraphics{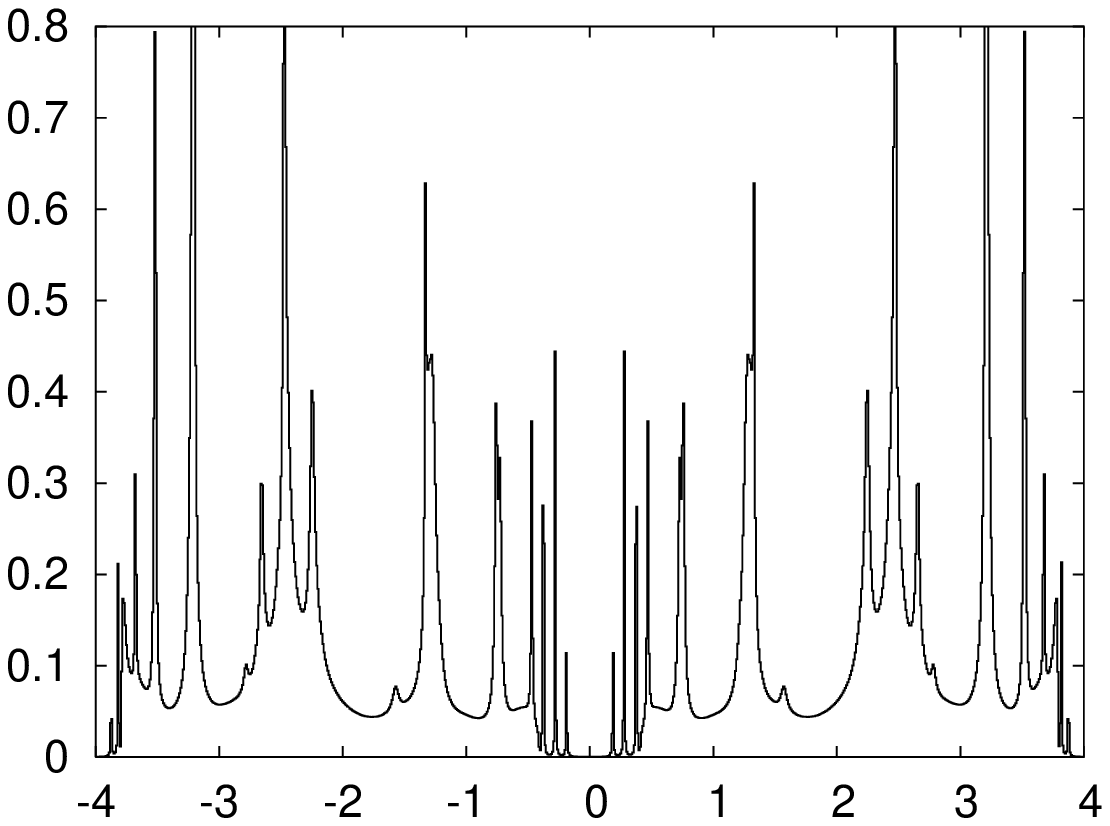}}\\
   \psfrag{MCPA}[B1][B1][2][0]{MCPA}\scalebox{0.33}{\includegraphics{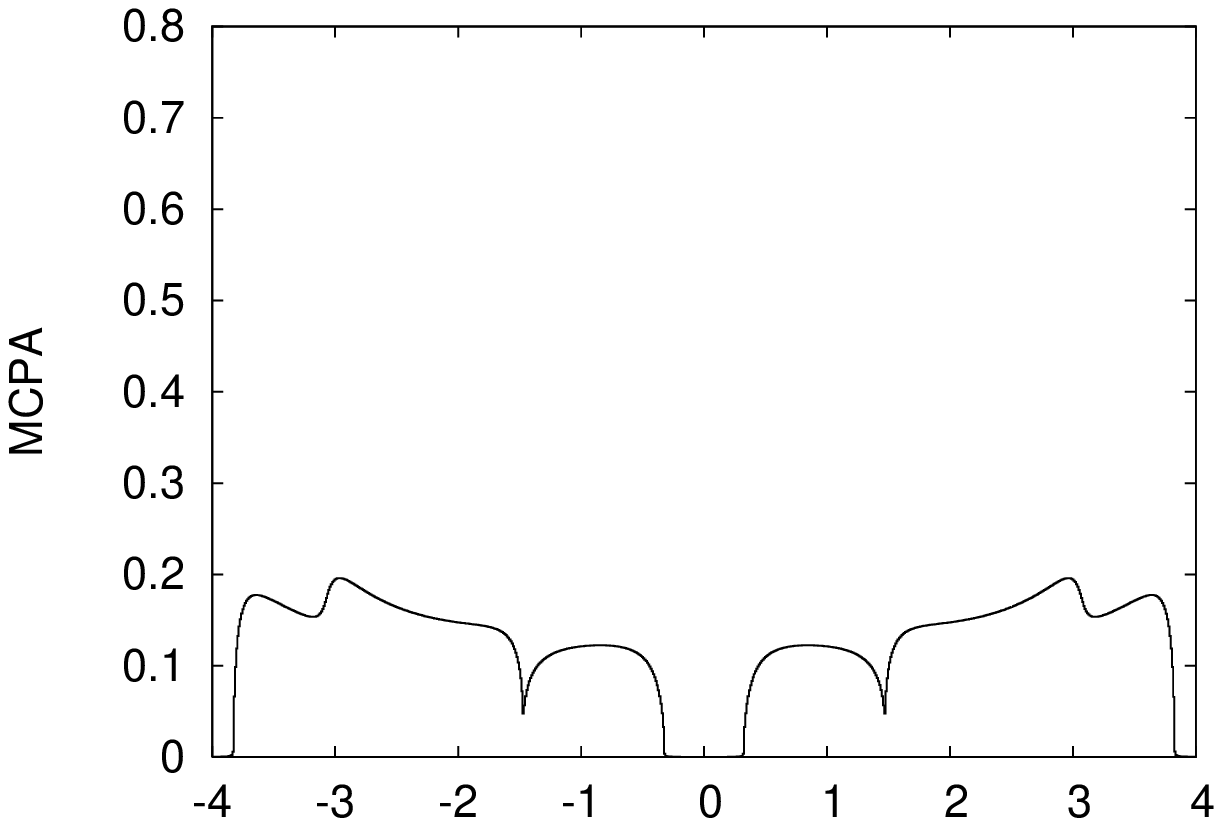}}
 & \scalebox{0.33}{\includegraphics{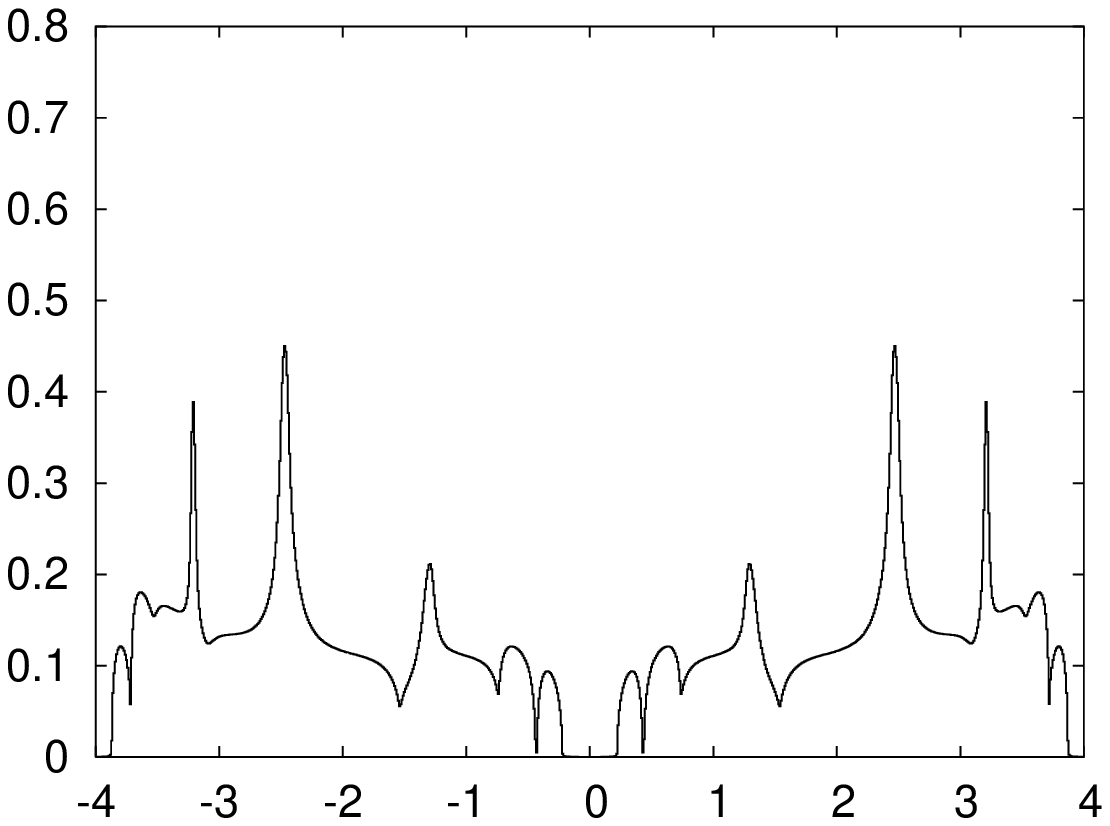}}
 & \scalebox{0.33}{\includegraphics{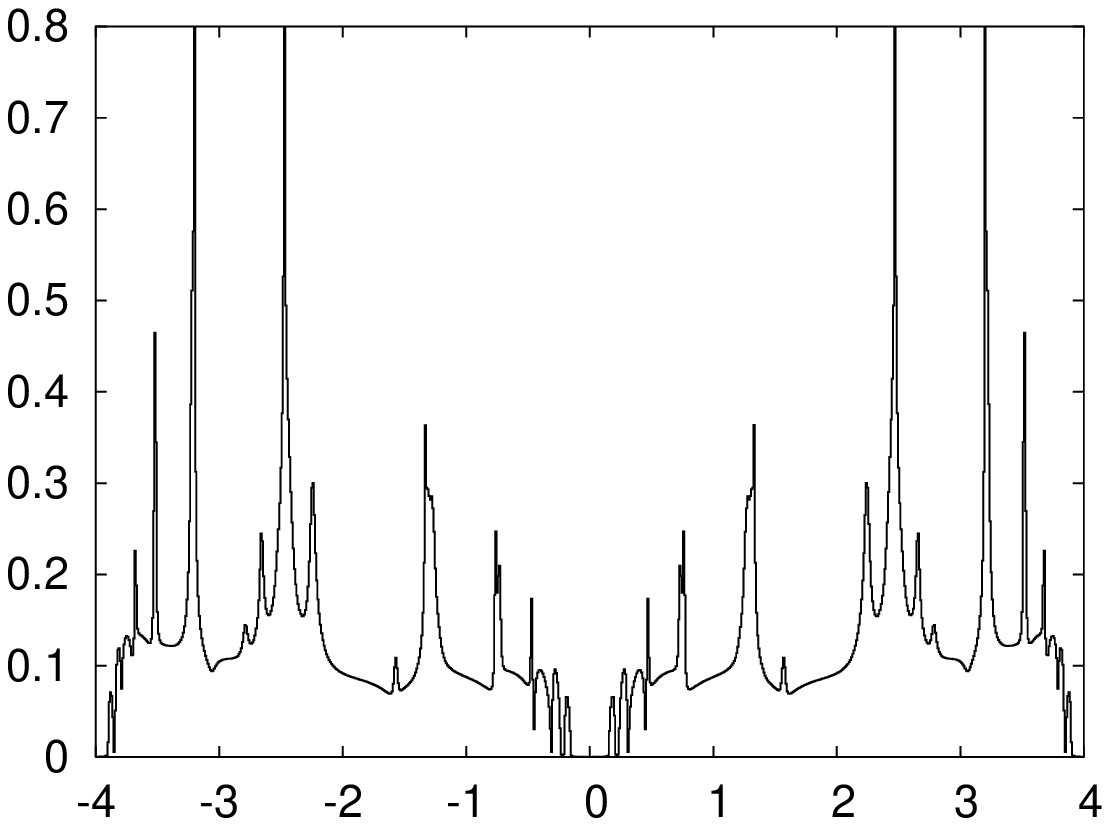}}\\
   \psfrag{NLCPA}[B1][B1][2][0]{NLCPA periodic}\scalebox{0.33}{\includegraphics{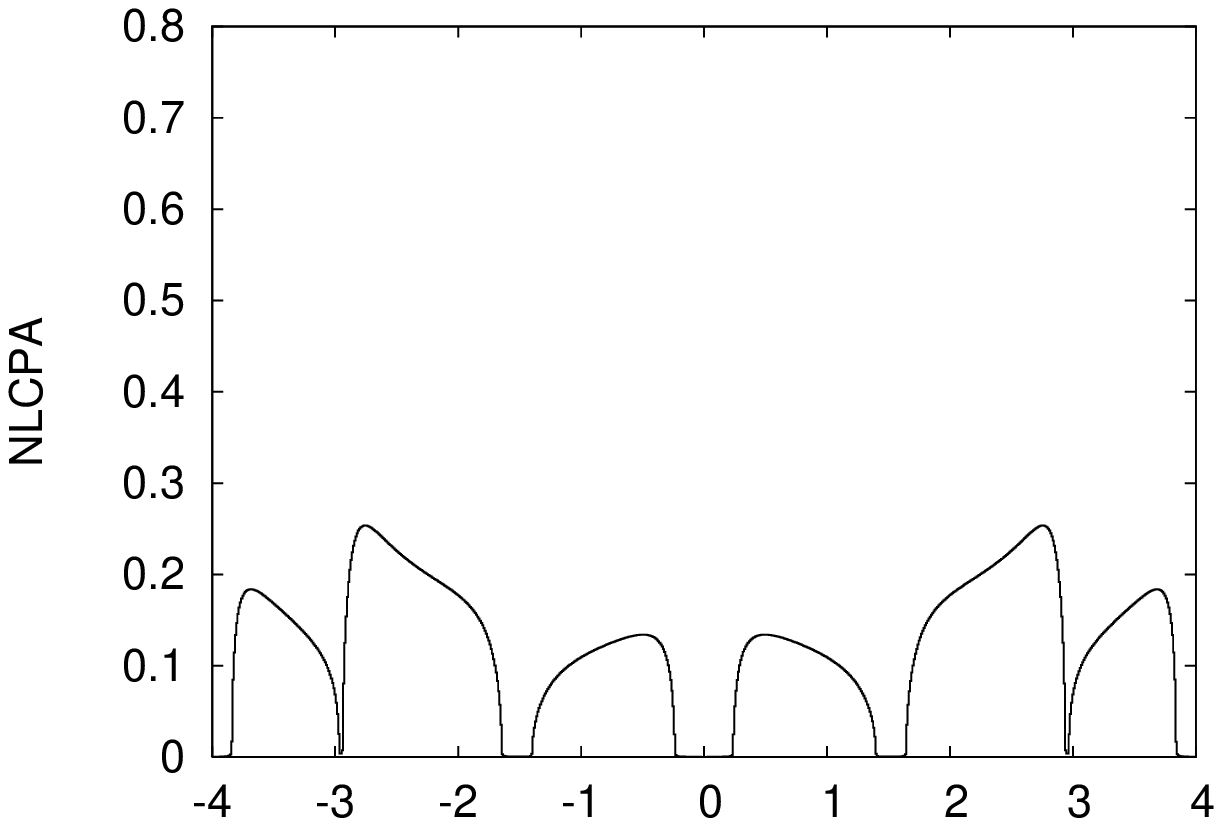}}
 & \scalebox{0.33}{\includegraphics{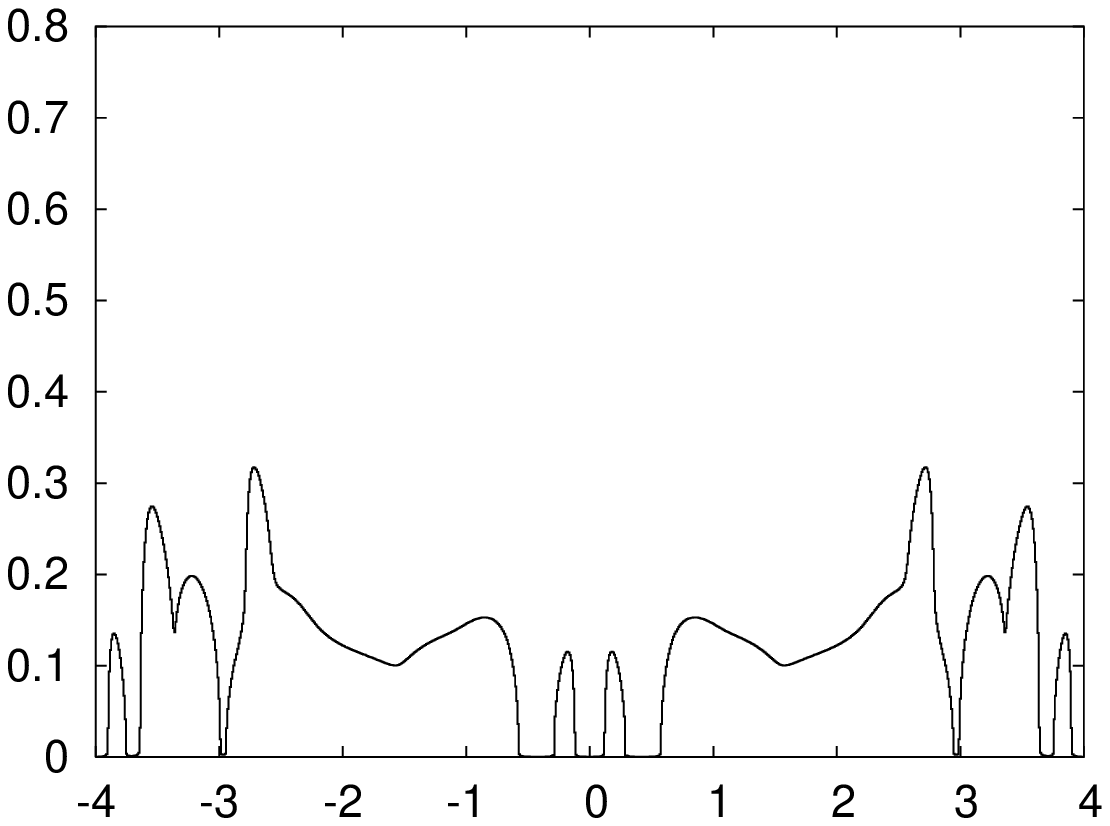}}
 & \scalebox{0.33}{\includegraphics{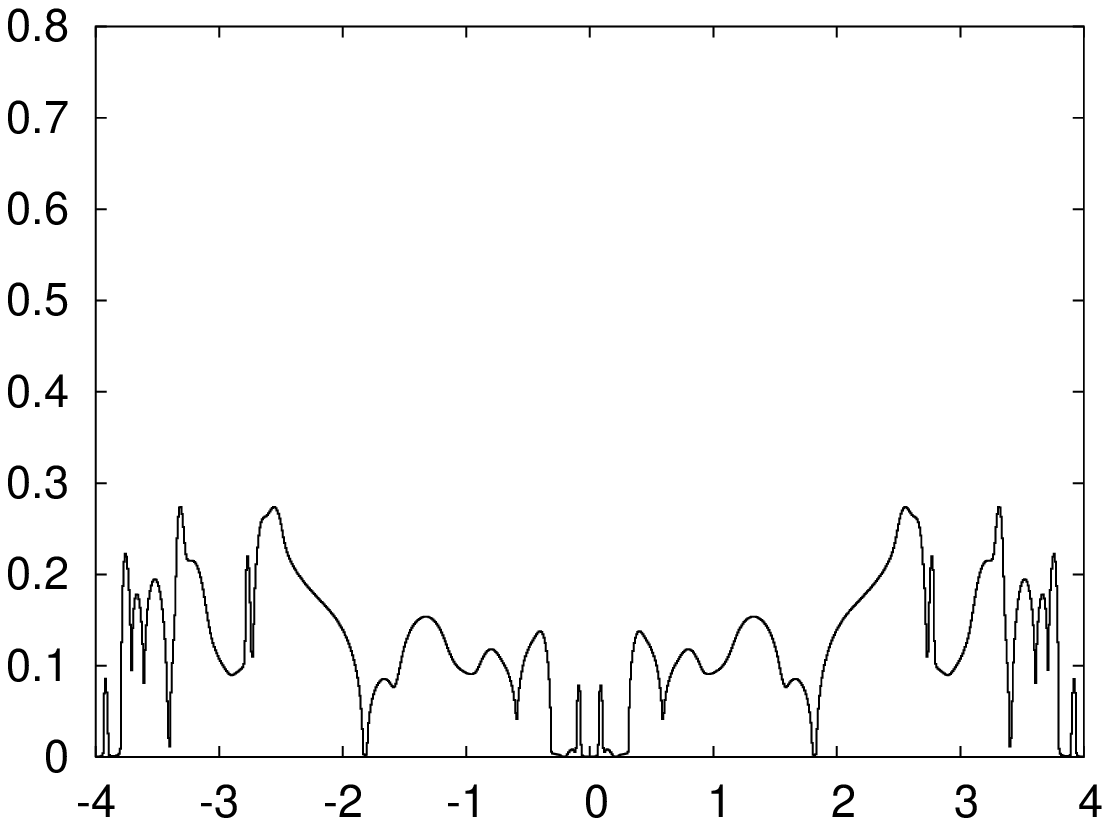}}\\
   \psfrag{NLCPA}[B1][B1][2][0]{NLCPA anti-periodic}\scalebox{0.33}{\includegraphics{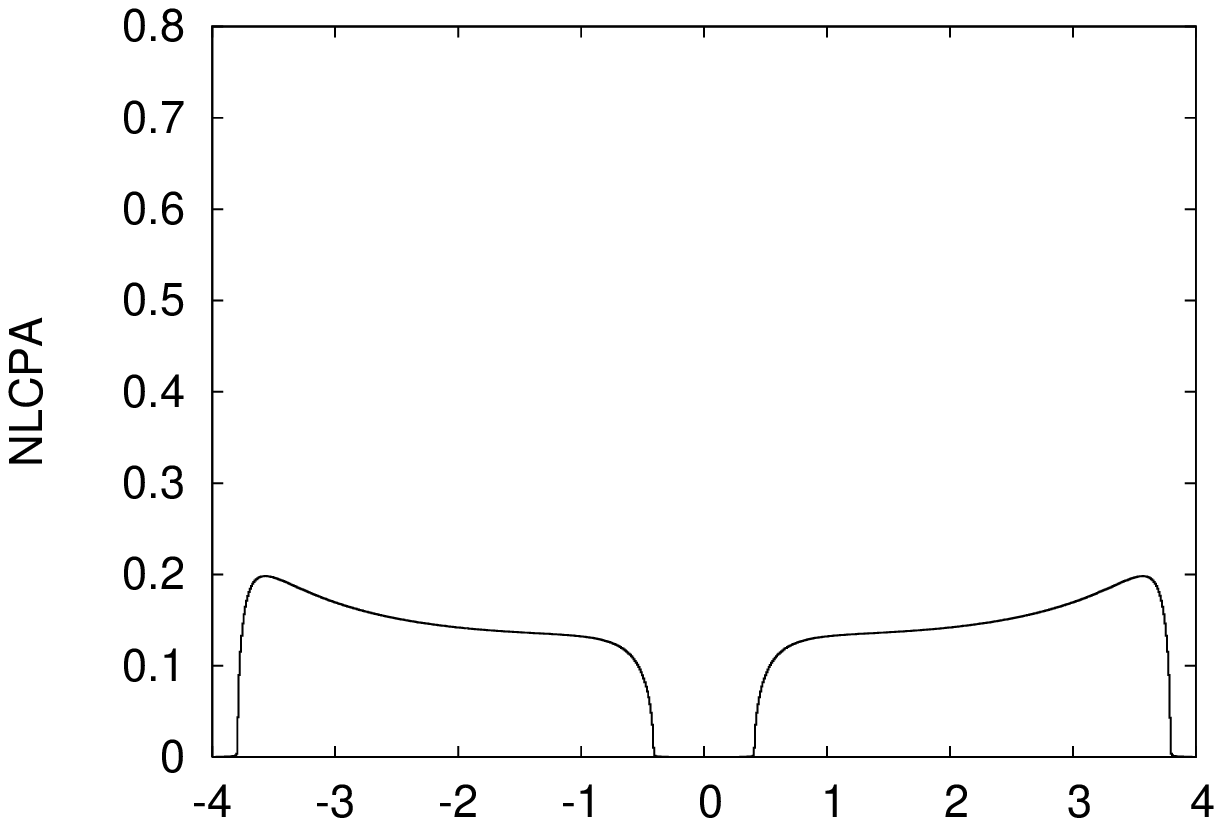}}
 & \scalebox{0.33}{\includegraphics{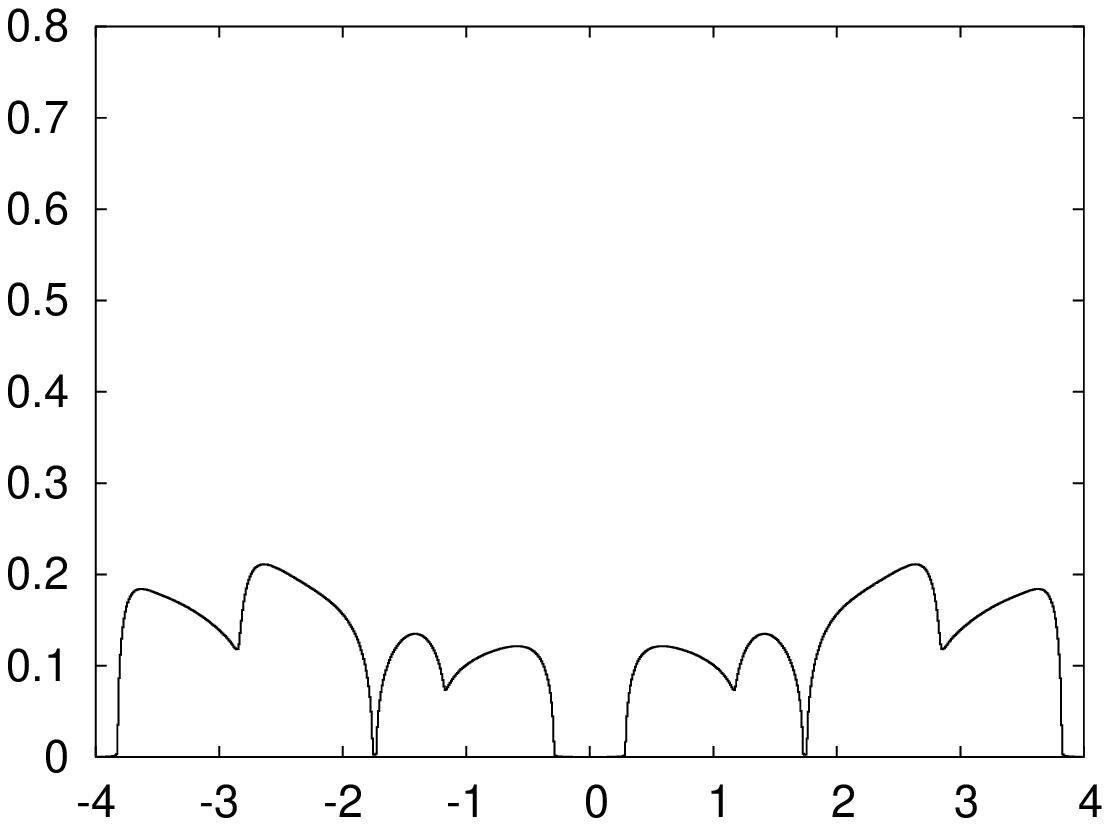}}
 & \scalebox{0.33}{\includegraphics{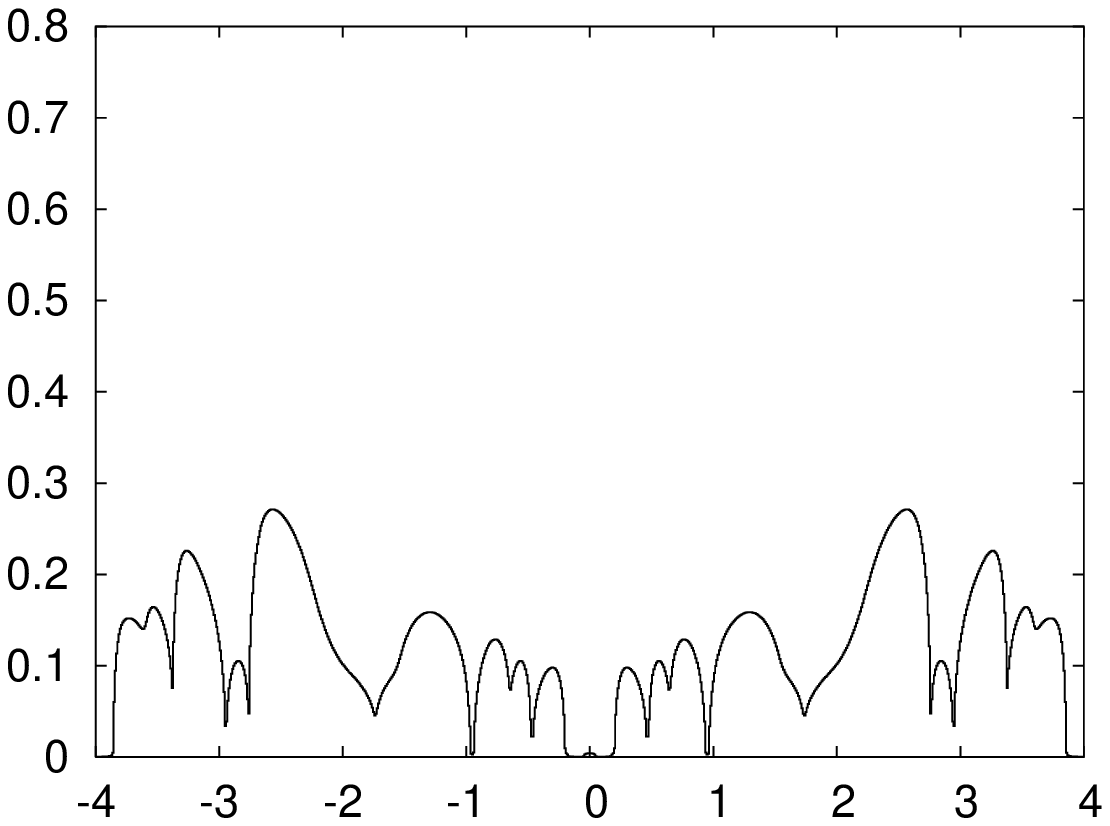}}\\
   \psfrag{Nc2}[B1][B1][2.5][0]{$N_C=2$}\psfrag{NLCPA}[B1][B1][2][0]{NLCPA mixed}\scalebox{0.33}{\includegraphics{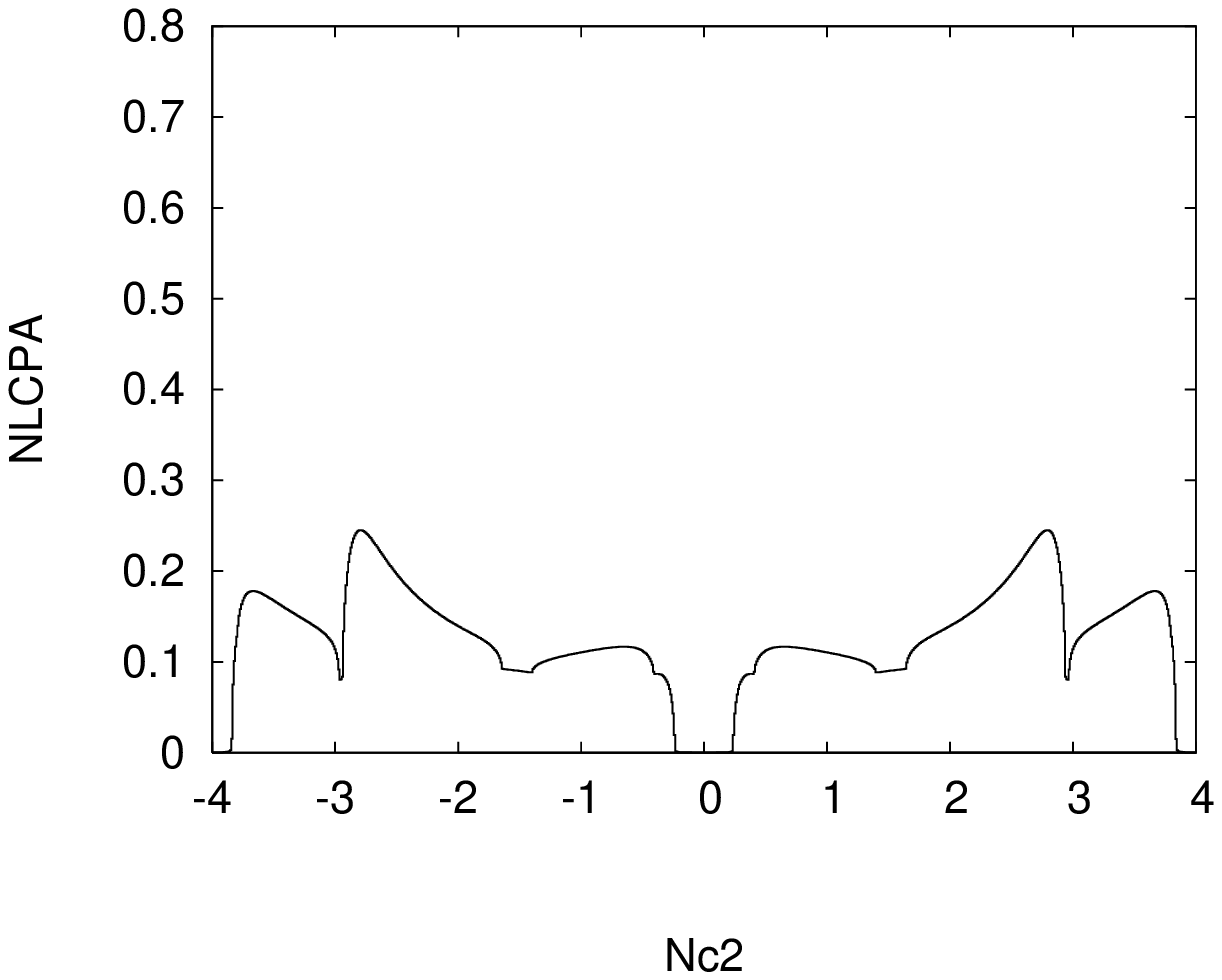}}
 & \psfrag{Nc4}[B1][B1][2.5][0]{$N_C=4$}\scalebox{0.33}{\includegraphics{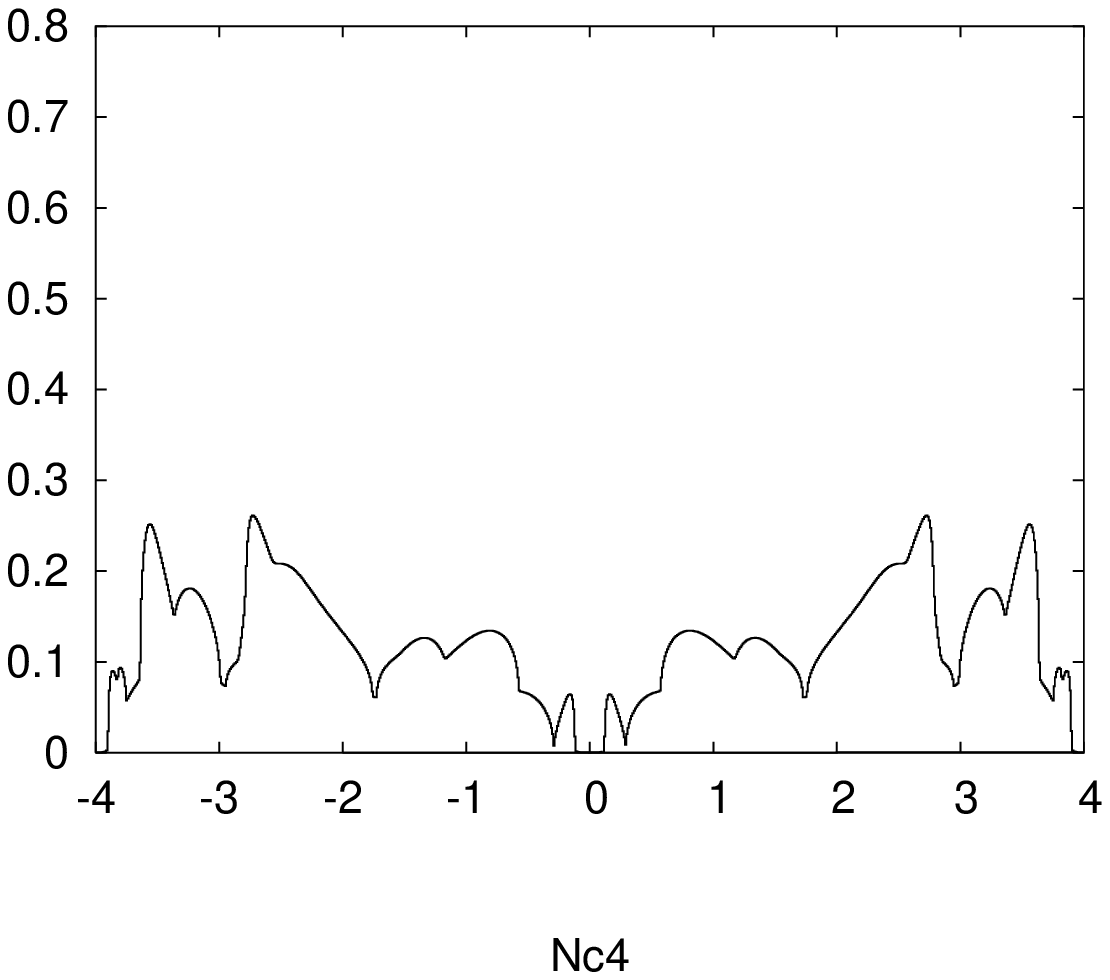}}
 & \psfrag{Nc6}[B1][B1][2.5][0]{$N_C=6$}\scalebox{0.33}{\includegraphics{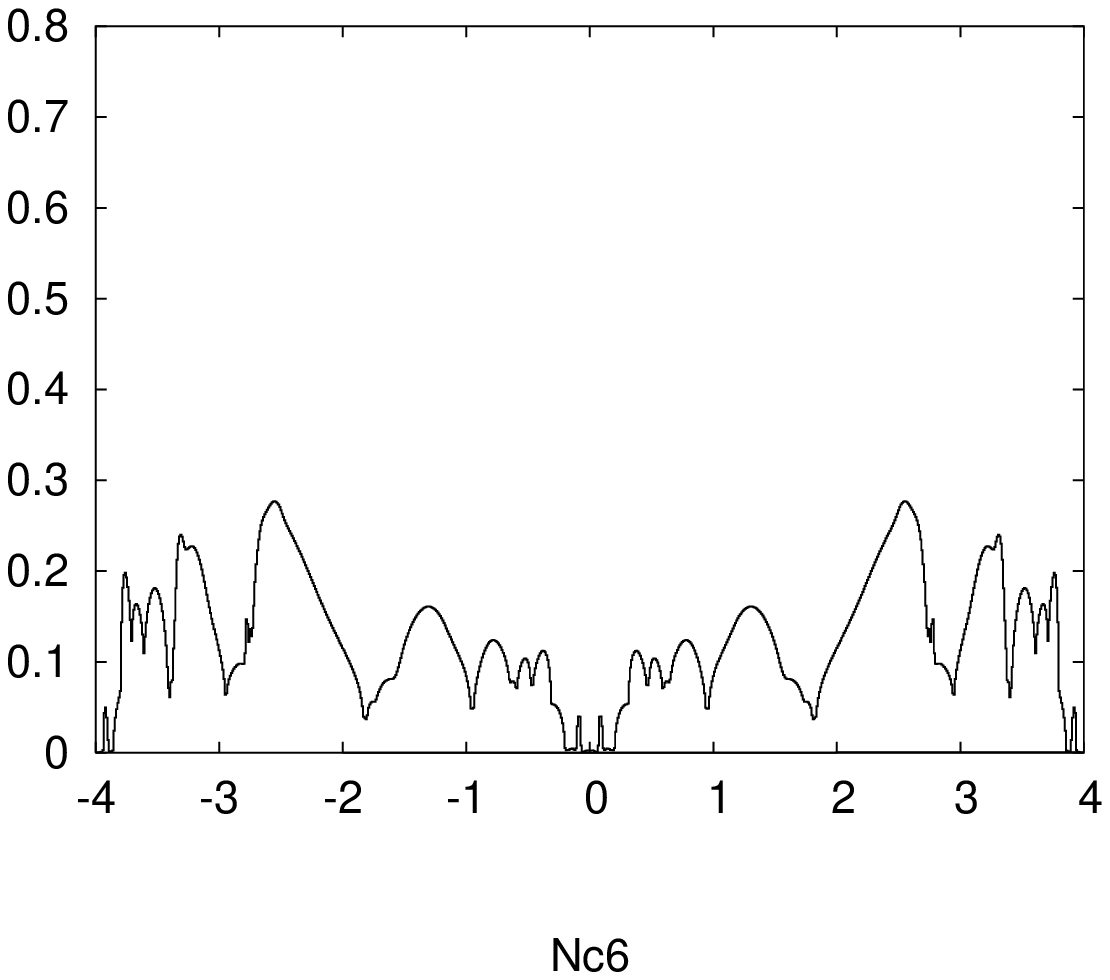}}
 \end{tabular}
 \caption{Configurationally-averaged density of states (DOS) per site as a function of energy $E$ for various cluster theories (top
 to bottom) with cluster sizes $N_c=2,4,6$ $(W=1,V_i=\pm 2)$ (left to right).}\label{cluster246}
 \end{center}
\end{figure}

The ideal test case for the validity of the new method is the simple 1D tight-binding model investigated in \cite{Rowlands2}. This is
because in 1D the exact result can be obtained numerically. Also, fluctuations are much more significant in 1D and so detailed structure is
expected in the DOS which can be accurately compared with both the exact result and other cluster theories such as the embedded cluster
method (ECM)~\cite{Gonis1} and molecular coherent-potential approximation (MCPA)~\cite{Tsukada1}. Indeed, in the absence of chemical
short-range order the NLCPA results should be very similar to those of the ECM and MCPA since the effects of self-consistency are less
significant in the random case. Note that in 1D there are only two NLCPA boundary conditions that are possible (periodic and anti-periodic) and the
new mixed Green's function is given by the simple expression
\begin{eqnarray}
 \overline{G}_{M}(\b{k}) &=& \overline{G}_{P}(\b{k})\cos^2\left((L_x/2)k_x\right) + \overline{G}_{AP}(\b{k})\sin^2\left((L_x/2)k_x\right) \nonumber\\
                  &=& \overline{G}_{P}(\b{k})\cos^2\left((N_ca/2)\b{k}\right) + \overline{G}_{AP}(\b{k})\sin^2\left((N_ca/2)\b{k}\right)
\end{eqnarray}

Figure 3 in \cite{Rowlands2} shows the exact DOS plot for a random
$A_{50}B_{50}$ alloy of constituent site energies $V_A=+2.0$ and
$V_B=-2.0$ (or written more concisely $V_i=\pm2$) with hopping
parameter $W=1.0$, together with the conventional CPA result.
 Figure
\ref{cluster246} in this manuscript shows results for the same model
obtained using the ECM, MCPA (averaged over all cluster sites) and
the NLCPA for cluster sizes $N_c=2,4,6$. The NLCPA anti-periodic
$N_c=2$ result is the same as the CPA here due to symmetry. First
note the large difference between the periodic and anti-periodic
NLCPA results for these small cluster sizes. Significantly, both the
periodic and anti-periodic results only resemble the ECM and MCPA
over particular and different energy regions. Indeed, troughs are
present at certain energies which are not seen in either the ECM or
MCPA results or the exact result given in figure 3 of
\cite{Rowlands2}. These unphysical features are caused by the
discontinuities in the Green's function. On the other hand, the new
mixed NLCPA result reassuringly looks remarkably similar to both the
ECM and MCPA results for all cluster sizes, with unphysical features
removed and troughs and peaks in the DOS which can be associated
with specific cluster disorder configurations. Note that all NLCPA
results become increasingly similar as the cluster size increases
and eventually converge to the same result at the critical cluster
size of $N_c=12$ \cite{Rowlands2}.

\subsection{Fermi surface of a two-dimensional model alloy}\label{fermi2d}

\begin{figure}[!]
\begin{center}
\subfigure{\psfrag{E}[B1][B1][1.2][0]{$E$}\psfrag{D}[B1][B1][1.2][0]{$DOS$}
\psfrag{M}[B1][B1][1][0]{$M$}\psfrag{A}[B1][B1][1][0]{$AP$}
\psfrag{P}[B1][B1][1][0]{$P$}\psfrag{x}[B1][B1][0.8][0]{$P_{x(y)}AP_{y(x)}$}
\scalebox{0.65}{\includegraphics[clip,type=eps,ext=.eps,read=.eps]{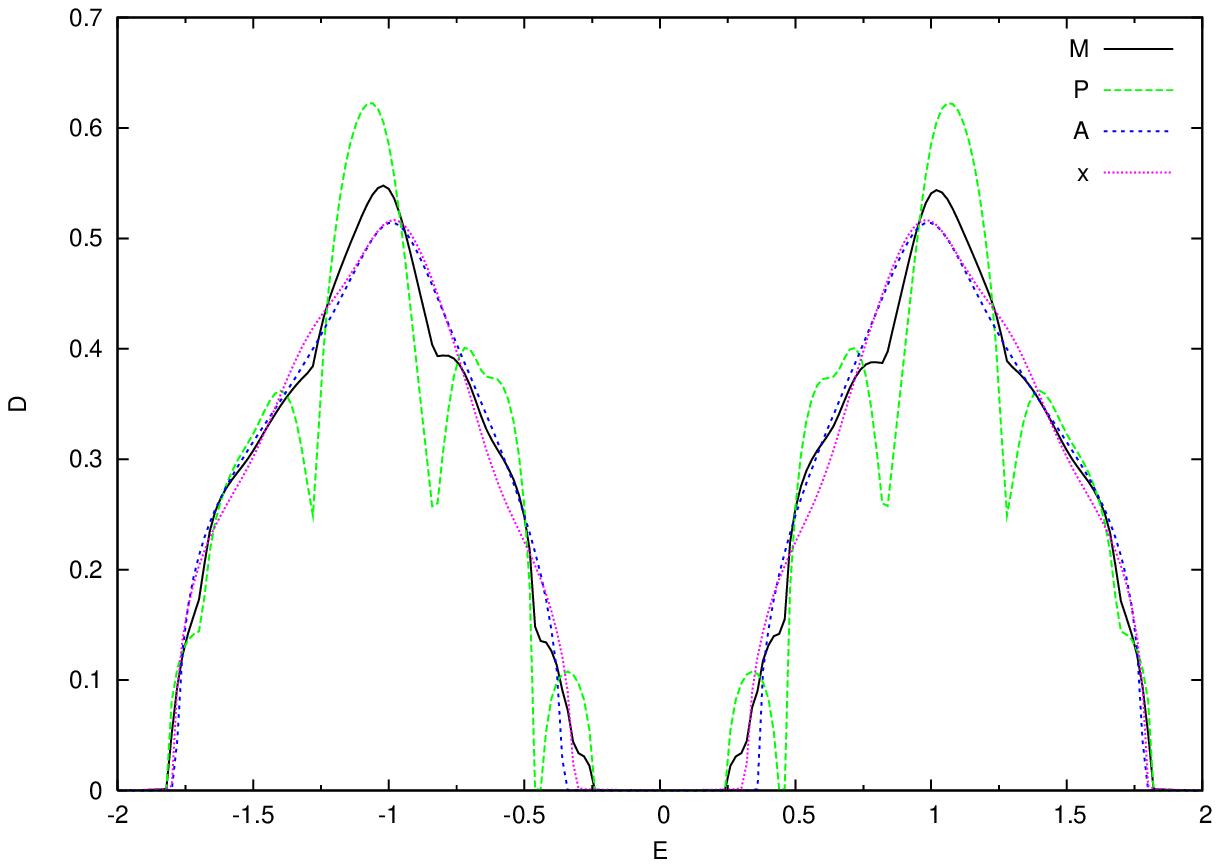}}} \\
\subfigure{\psfrag{M}[B1][B1][1][0]{$M$}\psfrag{P}[B1][B1][1][0]{$P$}
\psfrag{A}[B1][B1][1][0]{$AP$}\psfrag{y}[B1][B1][1][0]{$P_{x}AP_{y}$}
\psfrag{k}[B1][B1][1.2][0]{$((k_xa/\pi)^2+(k_ya/\pi)^2)^{1/2}$}\psfrag{x}[B1][B1][1][0]{$P_{y}AP_{x}$}
\psfrag{b}[B1][B1][1.2][0]{$A_{B}$}\psfrag{c}[B1][B1][1][0]{$X$}\psfrag{g}[B1][B1][1][0]{$\Gamma$}
\scalebox{0.65}{\includegraphics[clip,type=eps,ext=.eps,read=.eps]{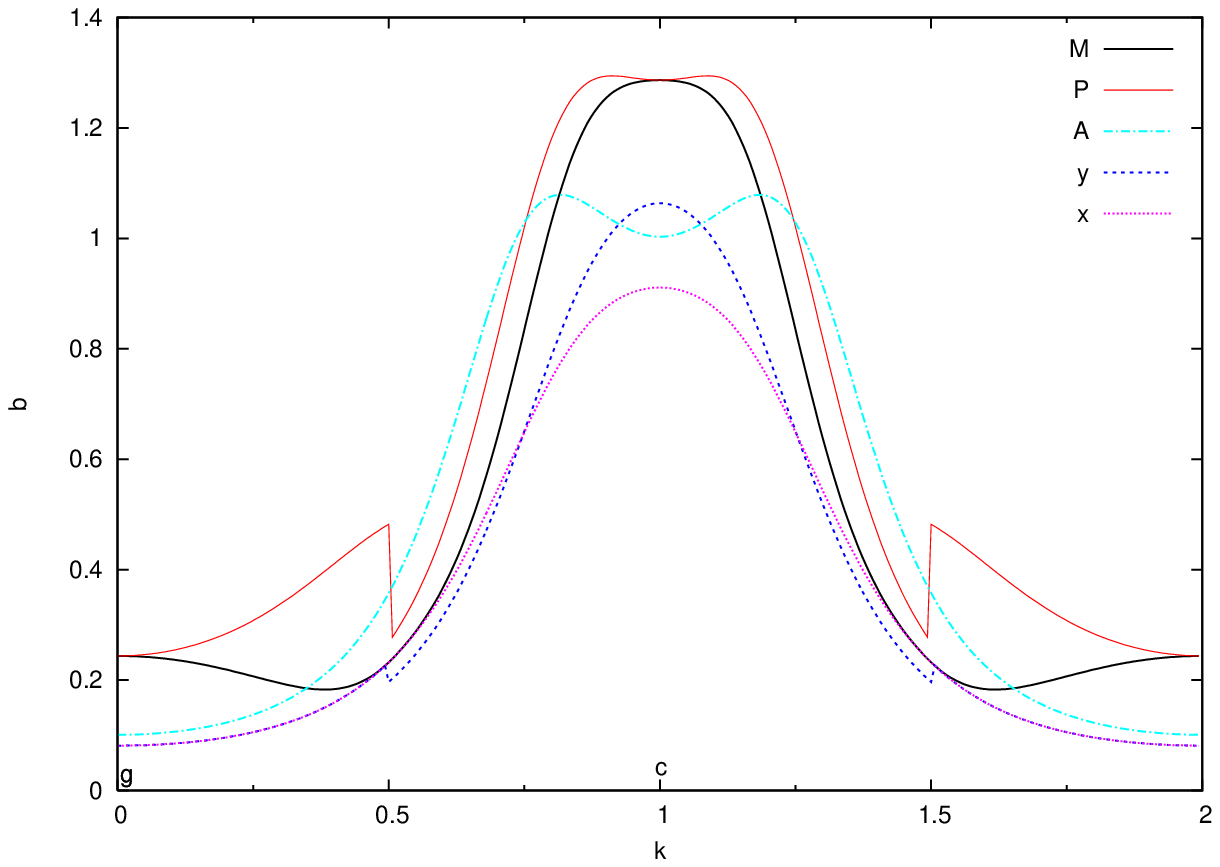}}} \\
\subfigure{\psfrag{M}[B1][B1][1][0]{$M$}\psfrag{P}[B1][B1][1][0]{$P$}
\psfrag{A}[B1][B1][1][0]{$AP$}\psfrag{y}[B1][B1][1][0]{$P_{x}AP_{y}$}
\psfrag{k}[B1][B1][1.2][0]{$((k_xa/\pi)^2+(k_ya/\pi)^2)^{1/2}$}\psfrag{x}[B1][B1][1][0]{$P_{y}AP_{x}$}
\psfrag{b}[B1][B1][1.2][0]{$A_{B}$}\psfrag{n}[B1][B1][1][0]{M}\psfrag{g}[B1][B1][1][0]{$\Gamma$}
\scalebox{0.65}{\includegraphics[clip,type=eps,ext=.eps,read=.eps]{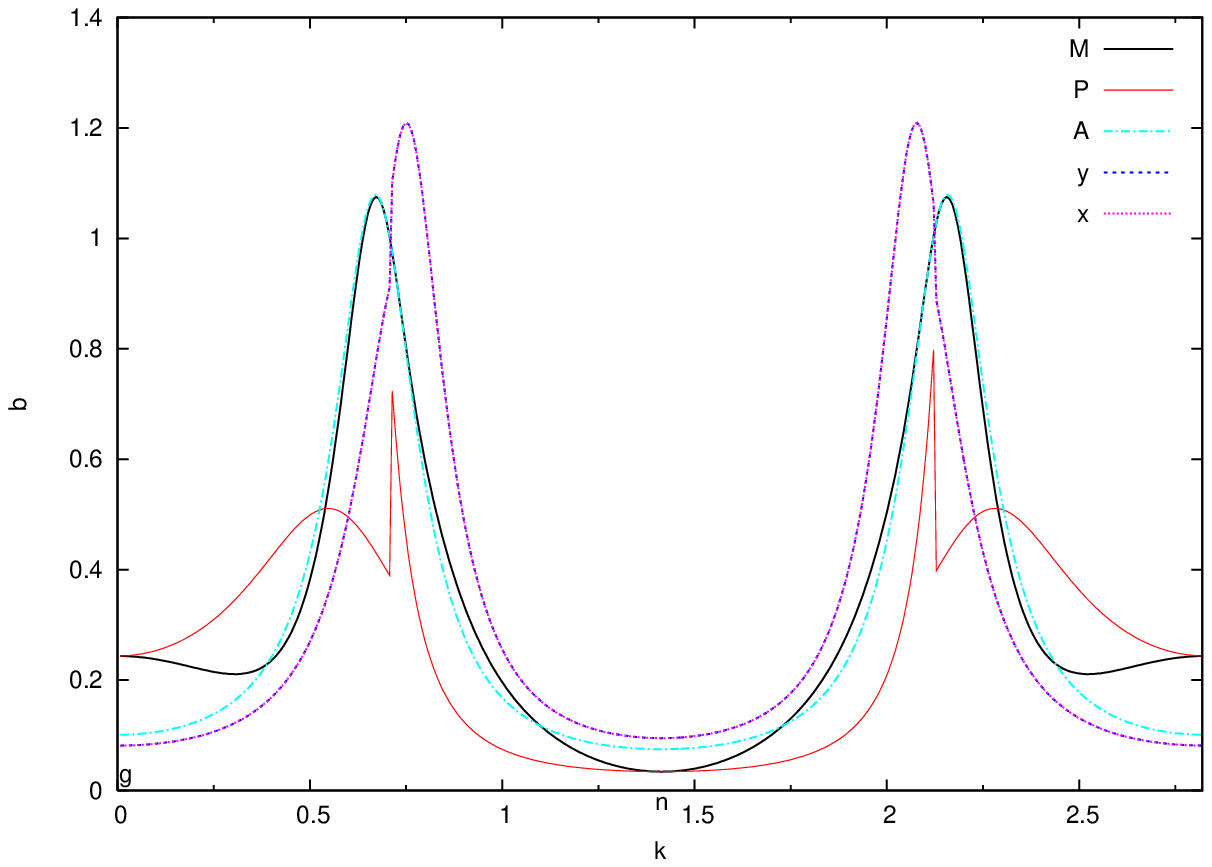}}}
\end{center}
 \caption{\label{dos} (colour online) \textbf{(top)} Split-band $N_c=4$ $(W=0.25,V_i=\pm 1)$ density of states (DOS)
 using periodic $(P)$, antiperiodic $(AP)$,
anti-periodic along x-axis $(P_y{AP}_x)$, anti-periodic along
y-axis$ (P_x{AP}_y$), and mixed (M) solutions, calculated from their
respective Green's functions. Where the DOS for $(P_y{AP}_x)$ and
$(P_x{AP}_y)$ are the same by symmetry.  To demonstrate the bounding
of the mixed Green's function in $\b{k}$-space, cross-sections of
$A_B(\b{k})$ at $E=-1$ are also plotted along the high symmetry axis
$\Gamma-{X}$ \textbf{(centre)} and $\Gamma-{M}$ \textbf{(bottom)}.
The mixed solution $A_B(\b{k})$ can be seen to remove
discontinuities introduced by the original NLCPA coarse graining
procedure at tile boundaries e.g $\pi/2a$. Any artefacts of the
original coarse graining are removed while correctly incorporating
the main spectral features of the separate $\overline{G}_i(\b{k})$.}
\end{figure}

\begin{figure}[!]
\begin{center}
\psfrag{C}[Bl][Bl][1.2][0]{CPA}\psfrag{N}[Bl][Bl][1.2][0]{NLCPA-Mixed}
\scalebox{0.6}{\includegraphics[type=eps,ext=.eps,read=.eps]{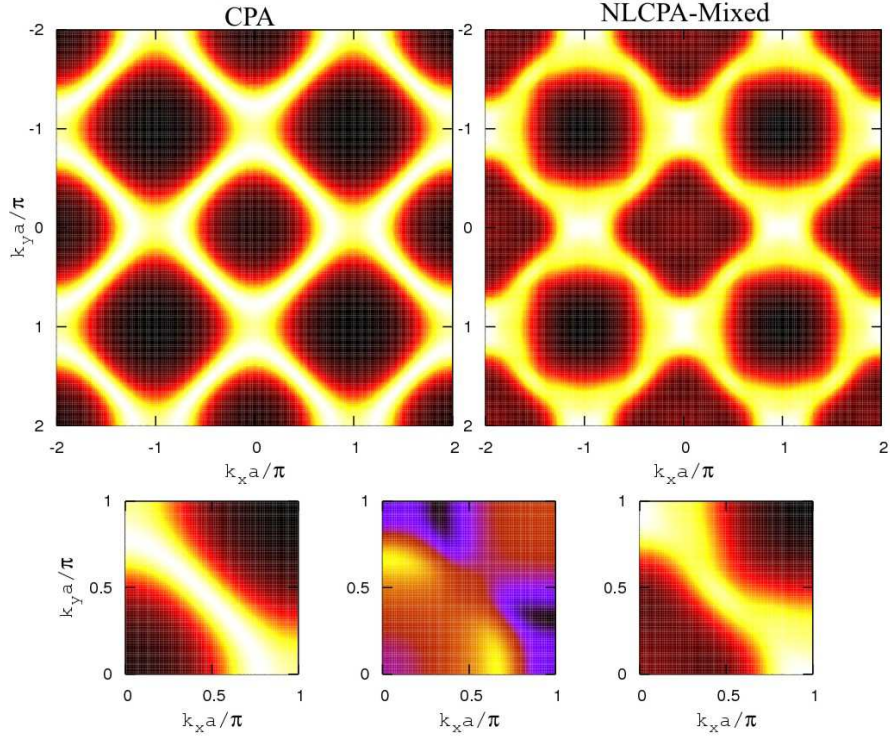}}
\caption{\label{spect} (colour online) $A_B(\b{k})$ at $E=-1.02$ and
$N_c=4$ $(W=0.25,V_i=\pm 1)$, plotted for the CPA
\textbf{(top-left)} and NLCPA-Mixed \textbf{(top-right)} in
$\b{k}$-space.
    Also shown is the plot zoomed in to the first quadrant of the Brillouin zone for CPA \textbf{(bottom-left)} and NLCPA-mixed \textbf{(bottom-right)} and the
    anisotropic difference $A_B^{CPA}(\b{k})-A_B^{M}(\b{k})$ \textbf{(bottom-centre)}. The colour scale for $A_B(\b{k})$ represents white as peaks going through yellow and red to troughs as
    black. The colour scale for the anisotropic difference $A_B^{CPA}(\b{k})-A_B^{M}(\b{k})$ represents positive peaks as yellow going through red and blue to negative peaks as black.}
\end{center}
\end{figure}

\begin{figure}[!]
\begin{center}
\subfigure{\psfrag{y}[B1][B1][1.2][0]{$DOS$}
\psfrag{x}[B1][B1][1.2][0]{$E$}\psfrag{a}[B1][B1][1][0]{$V_i\pm0.35$}
\psfrag{b}[B1][B1][1][0]{$V_i\pm
0.39$}\psfrag{c}[B1][B1][1][0]{$V_i\pm0.43$}
\psfrag{d}[B1][B1][1][0]{$V_i\pm0.45$}\psfrag{g}[B1][B1][1][0]{NLCPA-Mixed}
\psfrag{f}[B1][B1][1][0]{CPA}
\scalebox{0.65}{\includegraphics[clip,type=eps,ext=.eps,read=.eps]{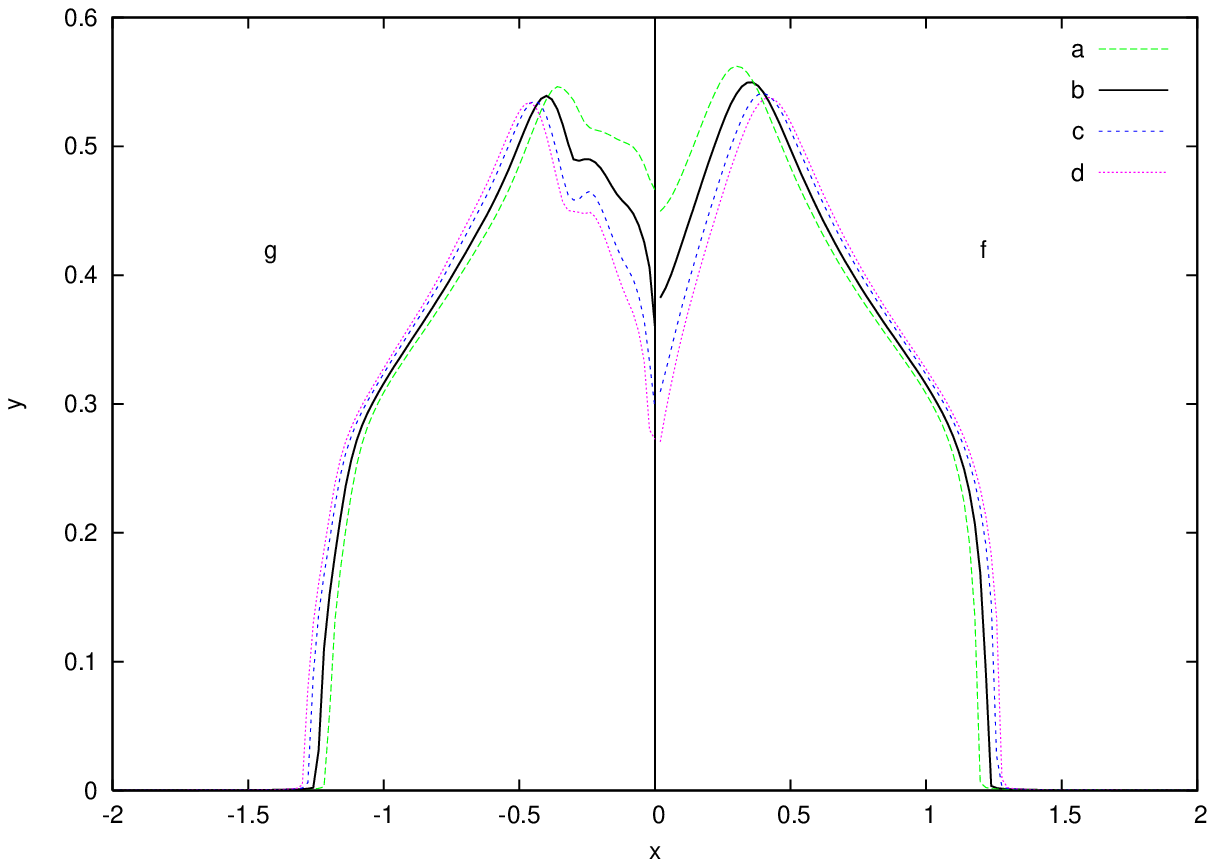}}} \\
\subfigure{\psfrag{a}[B1][B1][1][0]{$V_i\pm0.35$}
\psfrag{b}[B1][B1][1][0]{$V_i\pm0.39$}\psfrag{c}[B1][B1][1][0]{$V_i\pm0.43$}
\psfrag{d}[B1][B1][1][0]{$V_i\pm0.45$}\psfrag{N}[B1][B1][1][0]{NLCPA-Mixed}
\psfrag{C}[B1][B1][1][0]{CPA}
\scalebox{0.7}{\includegraphics[type=eps,ext=.eps,read=.eps]{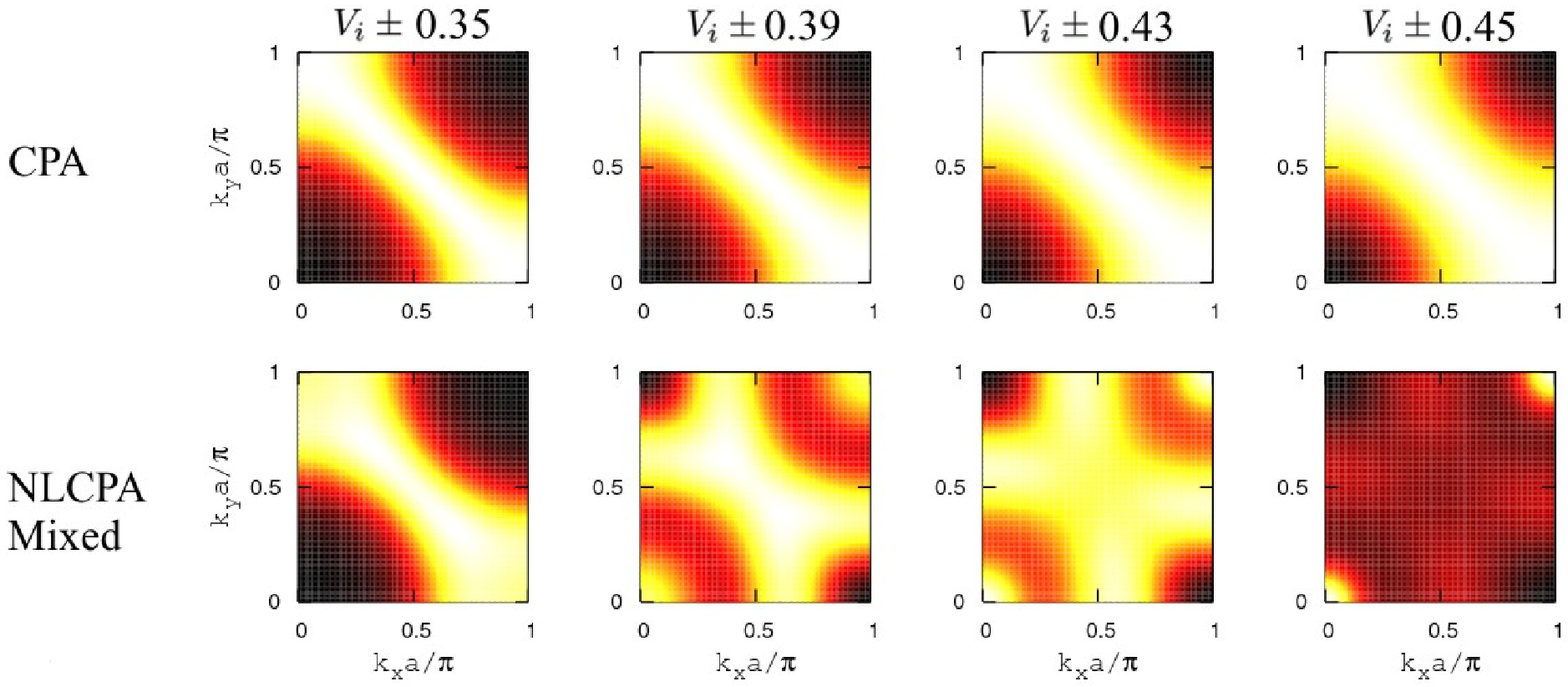}}}
\end{center}
\caption{\label{fermx} (colour online) \textbf{(top)} $N_c=4$
$(W=0.25)$ density of states (DOS) plotted for NLPCA-Mixed, and CPA
for varying atomic potentials $V_i$. Also shown \textbf{(bottom)}
are the Fermi energy plots of $A_B(\b{k})$ in the first quadrant of
the Brillouin zone for the half-filled band ($E=0$) for the CPA and
NLCPA-Mixed, where the colour scale represents white as peaks going
through yellow and red to troughs as black. As $V_i$ is increased a
significant change is observed between the CPA and NLCPA Fermi
surface topology. While the CPA retains a Fermi surface reminiscent
of a van Hove singularity, the mixed NLCPA produces distinct necking
features around the X-points. Looking at the DOS we see a small
fluctuation in the NLCPA-Mixed profile during the transition where
the DOS becomes lower than the CPA at the Fermi level.}
\end{figure}

In the 2D case of the NLCPA (and DCA) for systems with Green's
functions of the form (\ref{Gapprox}), spectral features are
unphysical for small cluster sizes due to the coarse graining
procedure. By having only a small number of tiles any large
discrepancy in the self-energy will result in a large discontinuity
at the tile boundaries and hence a Bloch spectral function that
remains essentially coarse grained in each tile. Furthermore, this
behaviour will be most evident near energies that represent
constituent ordered and clustered alloy states. Bulk quantities may
as a result display features that overcompensate for short range
correlations. For example in the DCA the stable d-wave state
observed for $N_c=4$ in the doped repulsive 2D Hubbard
model~\cite{Maier4} was later argued to be unfounded in comparison
to exact results~\cite{Su1}. The reason for this is that the $N_c=4$
periodic case exaggerates cluster induced d-wave behaviour by the
coarse graining procedure. Hence the DCA using periodic boundary
conditions can only be deemed to correctly predict such features for
large cluster sizes~\cite{Maier5}. It is very important from an
alloy point of view to improve upon the results available from small
periodic clusters, since a first principles treatment of large
clusters would require an unrealistic amount of computational power
(at the present time $N_c<4$ cluster sizes have been
investigated~\cite{Rowlands4,Rowlands5}). In this section we again
backup the validity of our method with a DOS comparison, then
investigate some of the new interesting Fermi surface physics
possible with the new implementation of the NLCPA. Also note that we
have used the concise convention of labelling the $V_A$ and $V_B$
atomic potential parameters by $V_i\pm V_A$ since we only use atomic
potentials symmetric about $E=0$ i.e. $V_B=-V_A$.

For this simple alloy model we investigate the first available
cluster after $N_c=1$ (CPA) i.e. $N_c=4$ with a nearest neighbour
hopping interaction $W=0.25$ and cluster principle vectors
$L_y=L_x=2a$. We maintain the same disorder and constituent alloy
concentrations as in the previous section for the 1d model i.e. a
random $A_{50}B_{50}$ alloy. The 4-site cluster is the ideal test
case since the difference between $\overline{G}_i(\b{k})$ should be
large and so any improvement given by $\overline{G}_M(\b{k})$ over
$\overline{G}_P(\b{k})$ should be clear. The interaction $W$ has
been given the same value as previous work~\cite{Jarrell1} in order
to compare with the DOS calculated there for the periodic $N_c=32$
case. For this lattice all solutions for $\overline{G}_i(\b{k})$ can
be considered almost indistinguishable from each other at
$N_c=32$~\cite{{Jarrell1,Hettler2}}, and hence \cite{Jarrell1}
provides a reliable guide to what the exact DOS should be. We first
examine the DOS (figure \ref{dos}) for the case of $V_i=\pm1$ where
the band has split and the nonlocal effects should be large, where
we will denote the periodic DOS by $DOS_P$ etc. As one might expect
we see that $DOS_P$ shows large fluctuations around $DOS_{CPA}$ (the
same as $DOS_{AP}$ by symmetry) that are unrealistic when compared
to previous work~\cite{Jarrell1}. Particulary the extra peak
formation is much stronger, and results in the DOS being far too low
at the troughs between peaks with 2 extra semi-bands being formed
near the main band gap. Indeed in the literature, for the $N_c=32$
case there is never such a large drop below the $DOS_{CPA}$. Looking
now at the new mixed result $DOS_M$ one sees that the large
oscillations associated with $DOS_P$ are no longer present but the
structure is much more like that of the $N_c=32$ case. Of note is
how the large troughs of $DOS_P$ are filled in: the DOS can be seen
to fill in the band gap, as well as the central peak becoming
prominent over the adjacent structure. In summary we see there is a
shifted and heightened peak at $E\approx\pm1.02$ and a second peak
emerging at $E\approx\pm0.79$, that overall corresponds to a clear
improvement over all $N_c=4$ solutions separately as well as the
CPA. This is more strong evidence that the mixing of NLCPA solutions
is physical for this model when carried out using the method
described in this paper.

The structure of $DOS_M$ is a consequence of how $A_B^M(\b{k})$
behaves in $\bf {k}$-space as shown in figure \ref{dos}. By
following each solution according to the weighting function
$\gamma_i(\b{k})$ the new solution $\overline{G}_M(\b{k})$ removes
discontinuities at all tile boundaries (irrespective of cluster
size) giving a smooth representation of the Bloch spectral function.
In figure \ref{dos} cross-sections of the spectral function
$A_B^M(\b{k})$ for an example energy show how the new solution is
bound and how it is an improvement. Where discontinuities are
present, the solution switches and uses a supposition of the ones
that remain analytic. For example $A_B^M(\b{k})$ at the
$\Gamma$-point must be equal to the periodic solution $A_B^P(\b{k})$
since all other solutions have a tile boundary at that point, and so
their associated $\{\gamma_i(\b{k})\}=0$.

A comparison can be made for the new spectral function
$A_B^M(\b{k})$ taking the Fermi energy at $E_F=-1.02$ (corresponding
to the main DOS peak of the mixed solution) with the CPA spectral
function for the same energy value, as shown in figure \ref{spect}.
Again, some quite remarkable features are apparent, most notably the
topological change induced by a kink halfway along the $X-X$
direction (not possible in the CPA) giving rise to a transition from
a circular-like surface to a diamond-like surface centred on the
$\Gamma$-point, and vice versa for the surface centred on the
$M$-point. This is quite dramatic since a Fermi surface (defined
spectral function peaks) taken at this energy would now have a whole
new set of nesting vectors such as the vector along $X-\Gamma -X$.
Moreover, these `rounding' and `squaring' features of Fermi surfaces
not seen in the CPA calculations have been observed experimentally.
For example, the work of Wilkinson \etal~\cite{Wilkinson1} gives a
clear example for $Cu_{1-x}Pd_x$ where the diamond-like Fermi
surface on the (100) plane predicted by the KKR-CPA is actually
found experimentally to be rounded. Furthermore, smearing and the
associated width of peaks regains a $\b{k}$-dependence that produces
so-called `hotspots' along the $\Gamma -X$ directions as shown by
dark patches in the difference between the CPA and mixed NLCPA. This
effectively produces necking in the surface at the $X$-point in the
direction $X-M$.

In the final set of results shown in figure \ref{fermx} we examine
the Fermi surface for the half-filled band with $E_F=0$. This is the
same for all $N_c=4$ NLCPA solutions and the CPA by symmetry. A
Fermi surface is only visible at this energy when there is finite
DOS and so we examine the properties where there is a breakdown of
Virtual Crystal (VCA) behaviour to a two-band system. This is
achieved by increasing the value of the atomic potentials $V_i$ from
zero. In the CPA one sees a transition from a band with a central
peak in the DOS, associated with a van Hove singularity at $E_F$, to
two peaks close to $V_i$ that represent the inclusion of states from
both constituent metals~\cite{Gonis1}. However, although this was a
great breakthrough at the time of the discovery of the CPA, no such
behaviour is observed in $\b{k}$-space. Sitting in the overlap of
metal bands one would expect the Fermi level to cross electron
states from both metals and hence produce a more complicated Fermi
surface. However, because the CPA has no $\b{k}$-dependence in
$\Sigma$ the correction to the VCA only emerges through an increased
smearing $\xi^{-1}\propto Im\{\Sigma(E_F=0)\}$. There is no
topological change in this system since $Re\{\Sigma(E_F=0)\}=0$.
This is because the alloy potentials are equidistant from zero and
so neither constituent has a preferential weighting. The CPA Fermi
surface is the same as the VCA Fermi surface at this energy and
alloy concentration i.e.~a straight line joining $X$-points. With
the inclusion of ${\bf {k}}$-dependence in the NLCPA a transition to
a more complicated form should be possible. However, associated with
such a transition (near a van Hove singularity) are very large
fluctuations in the correction to ${\bf {k}}$-space between coarse
grained tiles that are not identical. Hence, the standard (periodic)
solution to the NLCPA does not produce a physical result in this
regime. However, dramatically, we can now start to investigate such
phenomena using the new mixed solution to the NLCPA. Results are
plotted in figure \ref{fermx} for increasing $V_i$ and show how a
topological transition occurs at $V_i\approx\pm0.39$ from a CPA-like
Fermi surface. The Fermi surface splits by introducing new circular
surfaces around $M$ and $\Gamma$-points and also new necking
features around the $X$-points, in contrast to the CPA which only
displays increased smearing. This is very important because of the
new nesting vectors introduced~\cite{Varlamov1,Kordumova1}. The
emergence of this necking around $X$-points is very similar to that
observed in recent work for the shape-memory alloy
$Ni_{32}Al_{68}$~\cite{Dugdale1}. Comparison of first-principles
KKR-CPA calculations and experiment show the KKR-CPA does not
predict a necking feature in the Fermi surface at the $X$-points of
the Brillouin zone. The KKR-CPA retains a flat surface associated
with a van Hove singularity; the topological discrepancy is argued
to be the result of the Fermi level being close to a van Hove
singularity in the DOS. As $V_i$ is increased further the surface
around $M$ becomes prominent and may be associated with states on
the band edges of the pure $A$ and $B$ metal states emerging i.e
clustering of $A$'s or $B$'s, and it is believed $X$-point features
are associated with ordered alloy states. These points will be
further investigated in future work based on this paper~\cite{Batt1}
which will also incorporate chemical short-range order (which can be
introduced by weighting the cluster probabilities) and concentration
dependence $A_xB_{1-x}$.

\section{Conclusions}\label{conclusions}

This paper has presented a method for calculating the spectral function within the NLCPA. The method is general and can be applied to any
(allowed) cluster for any lattice in any dimension, and the obtained spectral function is fully $\b{k}$-dependent, continuous, and causal.
Furthermore, we believe the method to be the correct implementation of the NLCPA for cluster sizes below the critical cluster size (see
introduction) since it yields meaningful results for bulk properties such as the DOS and converges systematically as the cluster size
increases.

As an illustrative example, we have applied the method to a 2D tight-binding model alloy. Dramatically, a topological transition is clearly
observed in the Fermi surface as the difference in core potentials is varied in the split-band regime. Furthermore, Fermi surface splitting
has been observed producing new nesting features previously not observed in the conventional CPA. This occurs even in the absence of chemical
short-range order (SRO) which could be straightforwardly included by appropriately weighting the ensemble of cluster probabilities.

Clearly, it would be beneficial to apply the method within first-principles calculations. Indeed, a derivation and implementation of the
spectral function has recently been given within the first-principles KKR-NLCPA multiple-scattering theory in \cite{Tulip1}. However, only
the periodic set of cluster momenta $\{\b{K}_n\}$ are used and no attempt is made to remove the discontinuities. It is our intention to
combine the method in this paper with the formalism of \cite{Tulip1}, and investigate the effects of SRO on Fermi surface features of systems
such as $CuPd$~\cite{Wilkinson1}.

Finally, note that the one electron spectral function of correlated electron systems also have finite widths. Moreover, they can be measured
in angle-resolved photoemission experiments (ARPES). As they are due to dynamical fluctuations related to the many-electron system they can
be described by the DCA. Indeed, the DCA has recently been used to interpret such experiments~\cite{Maier3}. The method proposed here for
calculating the spectral function in the NLCPA can be directly transferred to the analogous calculation in the DCA and thus it may be useful
in this context as well as in the study of random alloys.

\ack

We thank B.~L.~Gy{\"{o}}rffy for useful discussions on Fermi surfaces. This work was funded by EPSRC~(UK).

\appendix

\section{Proof: causality and conservation of states}\label{causality}

It is straightforward to see that the new mixed Green's function
$\overline{G}_M(\b{k})$ is always causal. Provided the set of mixing
parameters satisfy (\ref{range}) and (\ref{Eqn:Sumgamma}), it
follows immediately from (\ref{mix}) that $\overline{G}_M(\b{k})$
will always be bound between the individual solutions
$\overline{G}_i(\b{k})$ in $\b{k}$-space. Since these individual
solutions have an imaginary part that is positive definite and hence
are causal~\cite{Jarrell1}, this implies that
$\overline{G}_M(\b{k})$ must also have an imaginary part which is
positive definite and hence is causal.

In order to prove that the mixing scheme preserves the total number
of electron states $N$, we introduce $\gamma_{i}(\b{k})$ as as a
coarse grained quantity in $\b{k}$-space $\gamma_{i}(\b{k})$ =
$\gamma_{i}({\bf {K_{\gamma}}})$, where $\{{\bf {K_{\gamma}}}\}$ are
an arbitrary set of $k$-points within the BZ that are each centred
on a coarse grained tile. If we also define $M_c$ as the total
number of tiles (independent of $N_c$), we have a step function
$0\leq\gamma_{i}({\bf {K_{\gamma}}})\leq1$ that has $M_c$ possible
values. For $M_c=1$ we have the simplest form of averaging with
$\gamma_{i}(\b{k})$ having no ${\bf {k}}$-dependence, equivalent to
a weighted average over the DOS. For $M_c>1$ coarse grained
$\b{k}$-dependence is introduced to the mixing of
$\overline{G}_i(\b{k})$, and $M_c\longrightarrow \infty$
representing the introduction of complete ${\bf k}$-dependent
mixing. We wish to show that conservation of electron states applies
to all values of $M_c$. To do this we integrate (\ref{mix}) over the
coarse grained $\gamma_i(\b{k})$ for an arbitrary $M_c$ to obtain
\begin{equation}\label{Eqn:reality}
 \frac{1}{\Omega_{\b{K}_\gamma}}\sum_{\b{K}_\gamma}\int \overline{G}_M(\b{k})d\b{k} =
 \sum\limits_i\gamma_i(\b{K}_\gamma)\overline{G}_i\left(\b{K}_\gamma\right)
\end{equation}
where the small ${\bf {k}}$-dependence has been integrated out,
leaving a new coarse grained version of $\overline{G}_M(\b{k})$,
$\overline{G}_M({\bf {K_\gamma}})$. Integrating (\ref{Eqn:reality})
over all energies $-\infty\leq E \leq\infty$, we obtain
\begin{equation}\label{Eqn:reality2}
 N_M(\b{K}_n)=\sum\limits_i\gamma_i(\b{K}_\gamma)\ N_i \left(\b{K}_n\right),
\end{equation}
where $N_M(\b{K}_n)$ and $N_i(\b{K}_n)$ are the total number of electron states for the mixed and $i$-th NLCPA solutions respectively for the
tile centred at ${\b{K}_\gamma}$. Note that this requires $\gamma(\b{K}_\gamma)$ to be energy-independent to keep $N_M$ in terms of $N_i$
i.e.~we do not want to infer any new self-consistency condition on $N_M$ to conserve $N$ apart from the NLCPA. Since each solution satisfies
the conservation laws separately and electron states are evenly distributed across the Brillouin zone, we have $N_i({\b{K}_n})=N/M_c$ for
each $i$. Now (\ref{Eqn:reality2}) becomes
\begin{equation}
 \frac{N_M}{M_c}=\frac{N}{M_c}\sum\limits_i\gamma_i(\b{K}_n)
\end{equation}
Provided (\ref{range}) and (\ref{Eqn:Sumgamma}) are satisfied so
that $\overline{G}_M(\b{k})$ is bounded between the individual
solutions at every point in $\b{k}$-space, we have for any $M_c$,
$N_M=N$. In particular we are free to use the $M_c\rightarrow\infty$
limit to introduce fully $\bf {k}$-dependent mixing of
$\overline{G}_i(\b{k})$.

\section*{References}

\bibliographystyle{unsrt}

\bibliography{gary}

\end{document}